# Real-space imaging of σ-hole by means of Kelvin probe force microscopy with subatomic resolution

B. Mallada[1,2,4]†, A. Gallardo[2,6]†, M. Lamanec[3,4]†, B. de la Torre[1,2], V. Špirko[3,7], P. Hobza[3,5]* and P. Jelinek[1,2]*

**Affiliations:**

[1]Regional Centre of Advanced Technologies and Materials, Czech Advanced Technology and Research Institute (CATRIN), Palacký University Olomouc, 78371 Olomouc, Czech Republic.

[2]Institute of Physics, Academy of Sciences of the Czech Republic, Prague, Czech Republic.

[3]Institute of Organic Chemistry and Biochemistry, Czech Academy of Sciences, Flemingovo Náměstí 542/2, 16000 Prague, Czech Republic

[4]Department of Physical Chemistry, Palacký University Olomouc, tr. 17. listopadu 12, 771 46, Olomouc, Czech Republic

[5]IT4Innovations, VŠB-Technical University of Ostrava, 17. listopadu 2172/15, 70800 Ostrava-Poruba, Czech Republic

[6]Department of Condensed Matter Physics, Faculty of Mathematics and Physics, Charles University, V Holešovičkách 2, 180 00 Prague, Czech Republic

[7]Department of Chemical Physics and Optics, Faculty of Mathematics and Physics, Charles University in Prague, Ke Karlovu 3, 12116 Prague, Czech Republic

† These authors contributed equally to this work and can be considered as the first author

*Corresponding author. Email: pavel.hobza@uochb.cas.cz, jelinekp@fzu.cz

**Abstract:** An anisotropic charge distribution on individual atoms, such as e.g. σ-hole, may strongly affect material and structural properties of systems. Nevertheless, subatomic resolution of such anisotropic charge distributions represents a long-standing experimental challenge. In particular, the existence of the σ-hole on halogen atoms has been demonstrated only indirectly through determination of crystal structures of organic molecules containing halogens or via theoretical calculations. Nevertheless, its direct experimental visualization has not been reported yet. Here we demonstrate that Kelvin probe force microscopy, with a properly functionalized probe, can reach subatomic resolution imaging the σ-hole or a quadrupolar charge of carbon monoxide molecule. This achievement opens new way to characterize biological and chemical systems where anisotropic atomic charges play decisive role.

**One-Sentence Summary:** We demonstrate that Kelvin force probe microscopy with properly functionalized tip can achieved subatomic resolution by real space imaging of the σ-hole.

**Main Text:**

**Introduction**

The observation of molecular structures with unusual atomic arrangement possessing two adjacent halogens or a pair of halogen atom and electron donor motifs (O, N, S, …) found in different crystals in the second half of the 20th century (*1–4*) represented a long-standing puzzle in supramolecular chemistry. Both halogens and electron donors are electronegative elements which carry a negative charge. Thus, close contacts of these atoms should cause highly repulsive electrostatic interaction. Counterintuitively, such atoms are frequently found to form intermolecular bonds, called latter halogen bond, which stabilizes the molecular crystal structure. Elegant solution offered by Auffinger et al. (*5*), Clark et al. (*6*), and Politzer et al. (*7, 8*) showed that formation of a covalent bond between a halogen and more electronegative atom (e.g. carbon) gives rise to so called σ-hole having anisotropic charge distribution on the halogen atom. Thus, a corresponding electrostatic potential, which is a physical observable, around halogen atom is not uniform (as considered within all empirical force fields) but exhibits, distal to covalently bound carbon, an electropositive crown surrounded by an electronegative belt, see **Figure 1A**.

Consequently, halogen bonding is attributed to attractive electrostatic interaction between a halogen's electropositive σ-hole and an electronegative belt of the other halogen or an electronegative atom with negative charge. The IUPAC definition of halogen bond (*9*) states that halogen bond "occurs when there is evidence of a net attractive interaction between an electrophilic region associated with a halogen atom in a molecular entity and a nucleophilic region in another, or the same, molecular entity". Stability of the σ-hole bonding is comparable to that of hydrogen-bonded complexes and also attraction in both types of noncovalent complexes was originally assigned to electrostatic interaction. While this is basically true for H-bonded complexes, in the case of halogen-bonded systems the importance of dispersion interaction (*10*) was highlighted. This is not surprising since in halogen bonded complexes close contact between two heavy atoms with high polarizability takes place.

The concept of halogen bonding was later generalized to σ-hole bonding and following the name of the electronegative atom bearing the positive σ-hole the halogen- (group 17), chalcogen- (group 16), pnicogen- (group 15), tetrel- (group 14) and, surprisingly, also aerogen-(group 18) bonding were recognized. The existence of a σ-hole in atoms of the mentioned groups of elements has common origin in the unequal occupation of valence orbitals.

The σ-hole bonding plays a key role in supramolecular chemistry (*11*) including engineering of molecular crystals or in biological macromolecular systems (*5*). Despite its relevance and an intensive research devoted to σ-hole bonding, the existence of σ-hole itself was confirmed only indirectly via quantum calculations (*5–8*) or crystal structures of complexes containing σ-hole donors and electron acceptors (*11–15*). But a direct visualization of this entity allowing to resolve its peculiar shape, has been missing so far.

Thus, we seek for a technique which imaging mechanism relies on the electrostatic force to facilitate the visualization of the anisotropic charge distribution on a halogen atom with the sub-Ångstrom spatial resolution. In the following, we show that real-space visualization of the σ-hole can be achieved by Kelvin probe force microscopy (KPFM) under ultra-high vacuum (UHV) conditions (*16*, *17*) with unprecedented subatomic resolution.

KPFM belongs to family of Scanning Probe Microscopy (SPM) techniques which routinely provides real space atomic resolution of surfaces. In the KPFM technique the variation of the frequency shift $\Delta f$ of oscillating probe on applied bias voltage V, which has quadratic form $\Delta f \sim V^2$ is recorded (*18*). The vertex of the Kelvin parabola $\Delta f(V)$ determines difference between work functions of tip and sample, also called contact potential difference $V_{CPD}$. Moreover, spatial variation of the contact potential difference $V_{CPD}$ across surface allows to map out local variation of surface dipole on the sample ($V_{LCPD}$) (*19*). Recent development of the KPFM technique operating in UHV conditions enabled to reach true atomic resolution on surfaces (*20*, *21*), imagine intramolecular charge distribution (*22*), control single electron charge states (*23*), bond polarity (*24*) or charge discrimination (*25*).

Atomic contrast in KPFM images originates from microscopic electrostatic force between static ($\rho_0$) and polarized charge densities ($\delta\rho$) located on frontier atoms of tip apex and sample when an external bias is applied (*19*) . There are two dominant components of such force including interaction between polarized charge on apex $\delta\rho_t$, which is linearly proportional to applied bias voltage (V), and static charge on sample $\rho_0$ and vice versa. Consequently, these two components cause local variation of the contact potential difference $V_{LCPD}$ (for detail description of the mechanism see SOM) providing atomic scale contrast.

Consequently, KPFM appears to be the tool of choice for imaging anisotropic charge distribution within a single atom, such as the σ-hole. To test this hypothesis, we deliberately choose tetrakis(4-bromophenyl) methane (4BrPhM) and tetrakis(4-fluorophenyl) methane (4FPhM) compounds (see **Figure 1A,B**), which skeleton arrangement facilitates a tripodal configuration once deposited onto a surface with a single bromine/fluor atom-oriented outwards of the surface (see **Figure S1**). This arrangement facilitates direct inspection of the σ-hole on a halogen atom by the front-most atom of a scanning probe, see **Figure 1C**. Deposition of the molecules in low coverage (less than 1 monolayer) on the Ag(111) surface held at room temperature under UHV conditions leads to a formation of well-ordered self-assembled molecular arrangements with rectangular fashion (see **Figures 2A,B**). Importantly, bromine atoms of 4BrPhM molecule have a significant positive σ-hole (see **Figure 1A**), while fluorine atoms possess isotropic negative charge (**Figure 1B**). This enables us to perform comparative measurements on very similar systems with/out the presence of σ-hole.

**Figures 2C,D** provide a comparison between 2D KPFM maps acquired over Br and F prominent atoms of the molecular compounds acquired with Xe decorated-tip, which reveals strikingly different contrast. In the case of the 4FPhM molecule we see a monotonous elliptical increase of the $V_{LCPD}$ signal over the fluorine atom. Differently, the KPFM image over the 4BrPhM molecule features a strike ring-like shape. It is worth noting that the 2D KPFM maps were recorded near the minimum of Δf-z curve to avoid effect of lateral bending of functionalized probe because of repulsive forces (*26*) causing image distortions.

To confirm the origin of the subatomic contrast observed experimentally on Br atom, we carried out KPFM simulations using static $\rho_0$ and polarized $\delta\rho$ charges of Br and F-terminated molecules and Xe-tip models obtained from Density Functional Theory (DFT) calculations (see **Figure S2**). **Figure 2E,F** show simulated KPFM images, which perfectly match to the experimental maps. Our theoretical model allows to decompose the two leading contributions including electrostatic interaction of polarized charge on tip $\delta\rho_t$ with static charge on molecule and counterpart term of the electrostatic interaction between polarized charge on molecule $\delta\rho_s$ with $\rho_t^0$ static charge of tip, see **Figure S2**. We found the subatomic contrast obtained on Br-terminated molecule can be rationalized from a variation of the microscopic electrostatic interaction between atomic scale charges of tip and sample. On the periphery of the Br atom, the positive shift of $V_{CPD}$ is given by the electrostatic interaction of the spherical polarized charge $\delta\rho_t$ of Xe-tip apex with the belt of negative charge surrounding the positive σ-hole. On the contrary, in the central part, the electrostatic interaction with positive crown of the σ-hole turned the $V_{LCPD}$ value with respect to that on the peripheral region. In the case of the 4FPhM molecule, both terms provide a trivial contrast with a positive shift of the $V_{LCPD}$ over the atom. Interestingly, the term corresponding to the static charge $\rho_s^0$ on the molecule reveals an elliptical shape originated from neighbor positively charged hydrogen atoms in underlying phenyl group of the 4FPhM molecule. Therefore, the shape of the feature presented in the KPFM image provides additional information about internal arrangement of the molecule on surface.

It is important to highlight that we deliberately employed single Xe atom to functionalize the tip-apex instead of more commonly used carbon monoxide (CO). As discussed above, Xe-tip allows us to optimize the imaging conditions of the σ-hole. The reason is that static charge density $\rho_0$ on apex of CO-tip has a strong quadrupolar character (see **Figure 3A**), while the charge on Xe-tip is highly spherical, see **Figure S2**. This choice eliminates spurious spatial variation of the $V_{LCPD}$ signal, which does not belong directly to the σ-hole. In particular, a component of the microscopic electrostatic interaction between static charge $\rho_t^0$ of tip and polarized charge on sample $\delta\rho_s$ needs to be abolished. In the case of Xe-tip, the spatial variation of the local $V_{CPD}$ is dominated by the component including interaction of a spherically polarized charge on Xe atom $\delta\rho_t$ with the anisotropic electrostatic field of the σ-hole. This enables a direct mapping of the spatial charge distribution of the σ-hole by means of KPFM technique

Thus, we consider instructive to look at the KPFM images acquired with CO-tip on the 4FPhM molecule as well. Despite the fact that the frontier fluorine atom of the 4FPhM molecule has isotropic charge distribution, experimental KPFM image shown on **Figure 3B** features a non-trivial ring-like shape with lower values of $V_{LCPD}$ signal on the center of the fluorine atom. Our KPFM simulation using CO-tip, shown on **Figure 3C**, coincides qualitatively with the experimental counterpart. We found, from detailed analysis of the electrostatic components (**see Figure S3**), that the subatomic contrast arises from the interaction of spherical polarized charge $\delta\rho_s$ on fluorine atom with the static quadrupole charge on CO-tip, composed of a negative crown of density on oxygen atom surrounded by a positive charge belt as shown on **Figure 3A**. Thus, the KPFM features resolved on the 4FPhM molecule reflects the quadrupolar charge distribution of the CO-tip. Thus, from spatial variation of the $V_{LCPD}$ signal, we can determine the sign of the quadrupole of the CO molecule on tip. The shift of $V_{LCPD}$ towards lower values in the central part of the KPFM image is caused by the negative charge crown of the quadrupole charge localized at oxygen, see **Figure 3A**. While the enhanced $V_{LCPD}$ value on the periphery reflects the positively charged belt of the quadrupole charge of CO molecule. This reverse shift of $V_{LCPD}$ with respect to the previous case of the σ-hole is caused by the fact that we inspect the anisotropic charge on the tip instead of sample. For a detailed explanation of the origin and sign of $V_{LCPD}$ shift, please see discussion in SOM.

Alternatively, some works reported subatomic features in non-contact Atomic Force Microscopy (nc-AFM) (*27*) images with CO functionalized tips (*28*). However, the origin of such contrast and their interpretation of the physical meaning are under debate (*29*, *30*). Additionally, nc-AFM has demonstrated unprecedented chemical resolution of single molecules (*31)* or their charge distribution (*32*). Thus, we consider appealing to explore the possibility to imagine the σ-hole by nc-AFM with functionalized tips (*28*).

**Figure S4** shows series of high-resolution nc-AFM images acquired at a wide range of tip-sample distances with CO-tip and Xe-tip, respectively. On the onset of atomic contrast in nc-AFM mode, the tip-sample interaction is dominated by an attractive dispersion. Resulting AFM contrast for both 4FPhM and 4BrPhM molecules has very similar spherical character lacking of any subatomic feature. Also, in close tip-sample distances, the AFM contrast remains very similar for both molecular compounds, featuring a bright spot in the center caused by the Pauli repulsion. Thus, we found that the AFM images do not reveal any signature of the σ-hole in whole range of tip-sample distances covering both attractive and repulsive interaction regime.

To understand in detail this experimental observation, we performed theoretical analysis of the nc-AFM images with CO-tip using the Probe Particle SPM model (*26*). **Figures S5** and **S6** display lateral cross-sections of different force components of the interaction energy acting between CO-tip and the outermost F and Br atoms of the 4FPhM and 4BrPhM molecules, respectively. Also, the calculated AFM images show very similar atomic contrast ruling out the possibility to image the σ-hole with CO-tip. From the analysis we inferred that the AFM contrast is dominated by dispersive and Pauli interaction, which both have highly spherical character. On the other hand, the electrostatic interaction possesses anisotropic character caused by presence of both σ-hole on Br atom and quadrupolar charge distribution on apex of CO-tip (see **Figure 3A**). Nevertheless, the magnitude of the electrostatic interaction is about one order smaller than the competing dispersion and Pauli interactions, which makes the σ-hole hard to image in the AFM technique. From this analysis we can conclude that the subatomic resolution of anisotropic charges requires a technique, such as KPFM, whose contrast mechanism is mastered by the electrostatic interaction mapping the charge distribution on forefront atoms.

In next, we investigated the influence of the σ-hole on the non-covalent intermolecular interaction energies. The nc-AFM technique provides the unique possibility to explore interaction energies between individual atoms/molecules placed on tip apex and sample via site-specific force spectroscopies (*33–36*). Apart of a quantitative evaluation of the interaction energies between well-defined entities, nc-AFM technique gives unvaluable opportunity to benchmark accuracy of the different theoretical methods to describe such weak non-covalent interactions (*35–37*).

The tip functionalization offers to explore distinct scenarios of the interaction mechanisms with molecular complexes. While Xe-tip has a positive net charge and large polarizability, CO-tip possesses a quadrupolar charge (O and C carry negative and positive net charge, respectively) and a relatively small polarizability. **Figure 4** displays their interaction energies with the 4FPhM and 4BrPhM molecules as function of the tip-sample distance. Small values of the maximum energies of 0.2 to 0.83 kcal/mol reveal non-covalent bonding mechanism. In general, the complexes with Xe-tip are more stable than these with CO-tip, which may be rationalized by larger dispersion interaction caused by Xe-tip. Interestingly, we observe that Xe-4BrPhM complex is less stable than the Xe-4FPhM complex (by 0.67 and 0.83 kcal/mol, respectively) despite the larger polarizability of Br determining the magnitude of the polarization interaction. This effect is caused by the presence of the repulsive electrostatic interaction between positive σ-hole on Br atom and positively charged Xe tip, which cancels partially the attractive dispersive interaction in the Xe-4BrPhM complex. On the other hand, the dispersive and electrostatic forces are both attractive in the case of the Xe-4FPhM complex resulting in larger total interaction energy. This observation not only support the presence of the positive σ-hole at Br atom, but it also underlines the origin of peculiar intermolecular orientation of halogen bonded molecular systems (*12–15*).

Recently, a vigorous effort has been devoted to the development of computational methods based on DFT with dispersion correction, being able to describe reliably intermolecular interactions in non-covalent complexes (*38*). But their transferability is still limited due to adopted approximations, and thus, careful benchmarking is desired. From this perspective, the above-described complexes represent interesting non-covalent systems for benchmarks with the complex interplay between the dispersion and the electrostatic interaction. It is worth noting that the maximum interaction energies measured are below 1 kcal/mol, which used to be considered as the limit of chemical accuracy strengthening further the benchmark.

It is important to remark that accurate interaction energies for different types of non-covalent complexes can be obtained from nonempirical coupled-cluster method covering triple-excitations (CCSD(T)). Unfortunately, its large computational demands make it impossible to apply this method to a system of the size of the molecules we investigated in the present work.
To circumvent this problem, we performed the CCSD(T) calculations on smaller reference model systems consisting of F- and Br-benzene, exhibiting similar characteristics as 4BrPhM and 4FPhM molecules (for details see SOM). We compared the calculated CCSD(T) interaction energies to interaction energies obtained with several popular DFT functionals, see **Table S3**. We found that range-separated ωB97X-V functional (*39*) covering implicitly dispersion energy provides very good agreement with the benchmarked data set as shown in **Table S3**. Since this functional was also shown (*39*) to provide best results among other popular DFT functionals for various types of systems with noncovalent interactions we selected this functional for the further use.

To check its transferability to our larger molecular systems, we calculated the interaction energies between 4FPhM and 4BrPhM molecules and Xe- and CO-tip models. **Figure 4** shows the excellent agreement between ωB97X-V interaction energies and experimental data resulted. The calculated energy minima for all complexes fit perfectly the measured values within the experimental error (see the inset of **Figure 4**). We should note, that PBE0 functional (*40*) with D3 correction (*41*) reproduce the CCSD(T) results on small model systems as well (see **Table S3** and **Figure S7**).

However, its transferability on the large systems was no longer as good as the range-separated ωB97X-V functional, see **Figure S8.**

The ωB97X-V functional describes very well the interaction trend for all considered systems (see inset of the **Figure 4**), with the fact that it systematically slightly overestimates the interaction energy by about a tenth of kcal/mol. The prefect agreement between theoretical and experimental values cannot be expected since calculations were limited to free standing 4FPhM and 4BrPhM molecules interacting with Xe- and CO-tip model while in experiment 4FPhM and 4BrPhM molecules were adsorbed at Ag(111) surface. The results confirm very good transferability of the ωB97X-V functional towards larger systems. Moreover, the very good agreement between calculated and experimental data sets obtained for all four complexes gives us also a confidence to the multi-scale benchmark technique employing small model complexes with the only Xe-tip model. Therefore, this approach makes it possible to accurately describe systems whose size does not allow the direct application of the accurate coupled-cluster technique (or similar) or direct experimental measurements are not feasible.

To conclude, we are convinced that the possibility to achieve the true subatomic resolution with KPFM technique not only provides the direct evidence of the existence of σ-hole but it substantially extends our possibilities to characterize charge distribution in complex molecular systems and on surfaces. We anticipate that this technique can be further extended to provide invaluable information about local inhomogeneous polarizability of individual atoms on surfaces or within molecules with the unprecedented spatial resolution in chemical and biologically relevant systems.

**References and Notes**

**Acknowledgments:** We acknowledge fruitful discussion with A. Růžička and P. Hapala. M. Lamanec acknowledges inspirative advices of his previous supervisor J. Kuchár. P.J. and B.T. dedicate this manuscript to memory of J.M. Gómez-Rodríguez.

**Funding:**

Czech Science Foundation  GACR 20-13692X (A.G., B.M., P.J.) 19-27454X (P.H.)

Praemium Academie of the Academy of Science of the Czech Republic (A.G.)

Palacký University Internal Grant Association IGA_PrF_2021_031 (M.L.)

Palacký University Internal Grant Association IGA_PrF_2021_034 (B.M.)

CzechNanoLab Research Infrastructure supported by MEYS CR (LM2018110).

**Author contributions:**

Conceptualization: P.J.

Methodology: B.M., A.G., B.T., P.J.

Theoretical calculations: A.G., M.L., V.Š., P.J., P.H.

Experimental: B.M., B.T.

Funding acquisition: P.H., P.J.

Supervision: B.T., P.H., P.J.

Writing – original draft: B.M., A.G., B.T., P.H., P.J.

**Competing interests:** Authors declare that they have no competing interests.

**Data and materials availability:** The data that support the findings of this study are available from the corresponding authors upon reasonable request.

Supplementary Materials

Materials and Methods

Supplementary Text

Figs. S1 to S17

Tables S1 to S3

References (*42–56*)

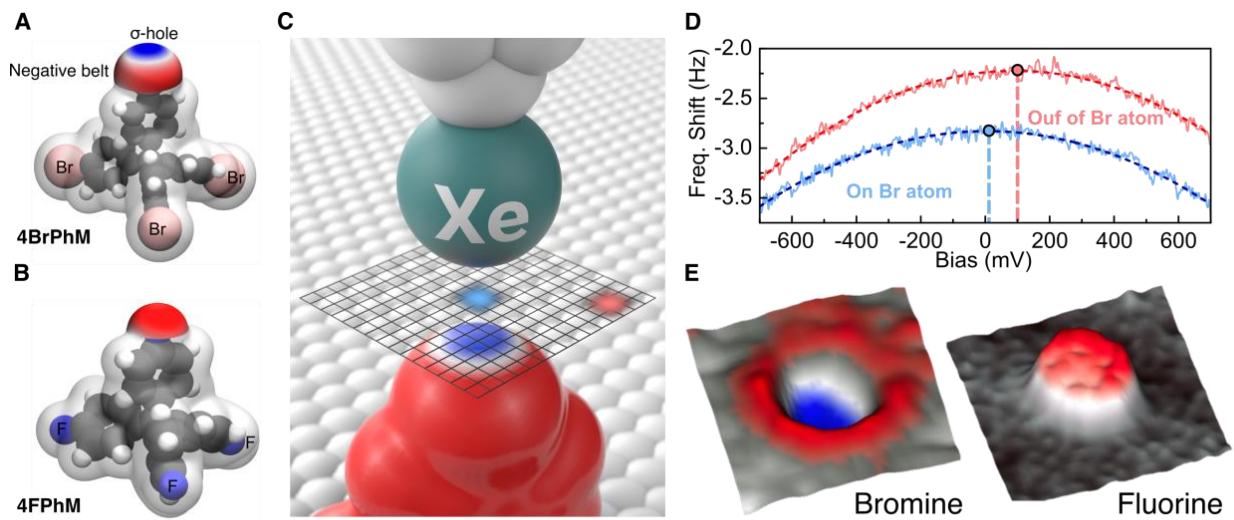

**Fig. 1. Schematic view of the KPFM measurements to image σ-hole. A-B** Models of 4BrPhM and 4FPhM molecules including corresponding electrostatic potential map on outermost Br/F atom. They reveal the presence of the σ-hole on Br atom, while there is isotropic negative charge on F atom. **C** Schematic view of acquisition method of the KPFM measurement with a functionalized Xe-tip on 2D grid. **D** corresponding Δ$f$(V) parabolas acquired in the central part (blue) of the 2D grid and on periphery (red). Vertical dashed lines indicate the value of $V_{LCPD}$ for the given Δ$f$(V) parabola, which forms 2D KPFM image. **E** 3D representation of the KPFM images ($V_{LCPD}$ maps) acquired with Xe-tip over a bromide and fluoride atoms of 4BrPhM and 4FPhM molecules (blue color represents low value of $V_{LCPD}$ and red color high value of $V_{LCPD}$, respectively).

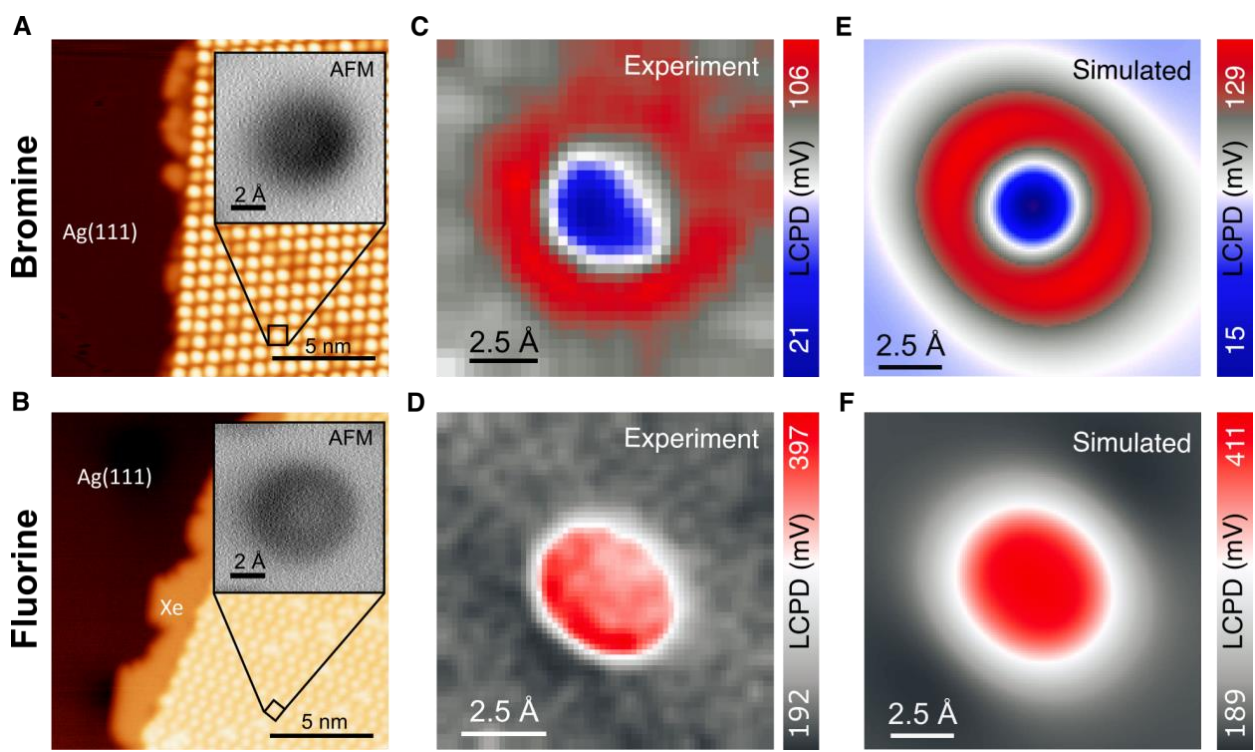

**Fig. 2. KPFM imaging of 4BrPhM and 4FPhM molecules with Xe-tip. A-B** STM images of molecular self-assembled sub-monolayer of 4BrPhM and 4FPhM molecules on Ag(111) surface. Inset images display AFM images acquired on a single molecule with Xe-tip at the minima of the frequency shift **C-D** Experimental KPFM images obtained with a functionalized Xe-tip over a bromide and fluoride atoms of a single 4BrPhM and 4FPhM molecules. **E-F** Calculated KPFM images with a functionalized Xe-tip of a single 4BrPhM and 4FPhM molecules. (blue color represents low value of $V_{LCPD}$ and red color high value of $V_{LCPD}$, respectively).

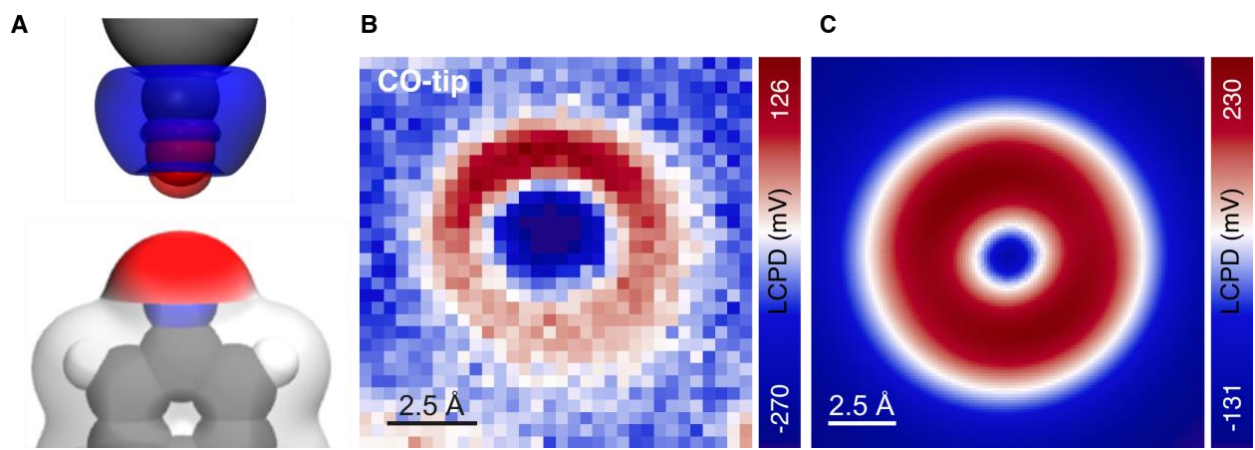

**Fig. 3. KPFM imaging of 4FPhM molecule with CO-tip. A** Schematic view of CO-tip above a 4FPhM molecule with superimposed calculated differential charge density of CO-tip revealing quadrupole charge of a CO-tip model (up) and calculated electrostatic potential of 4FPhM molecule showing an isotropic negative charge on the frontier fluorine atom in 4FPhM (down). **B.** Experimental KPFM image acquired over the frontier fluorine atom with a CO-tip. **C** Simulated

KPFM image of 4FPhM molecule with CO-tip. (blue color represents low values of $V_{LCPD}$ and red color high values of $V_{LCPD}$, respectively).

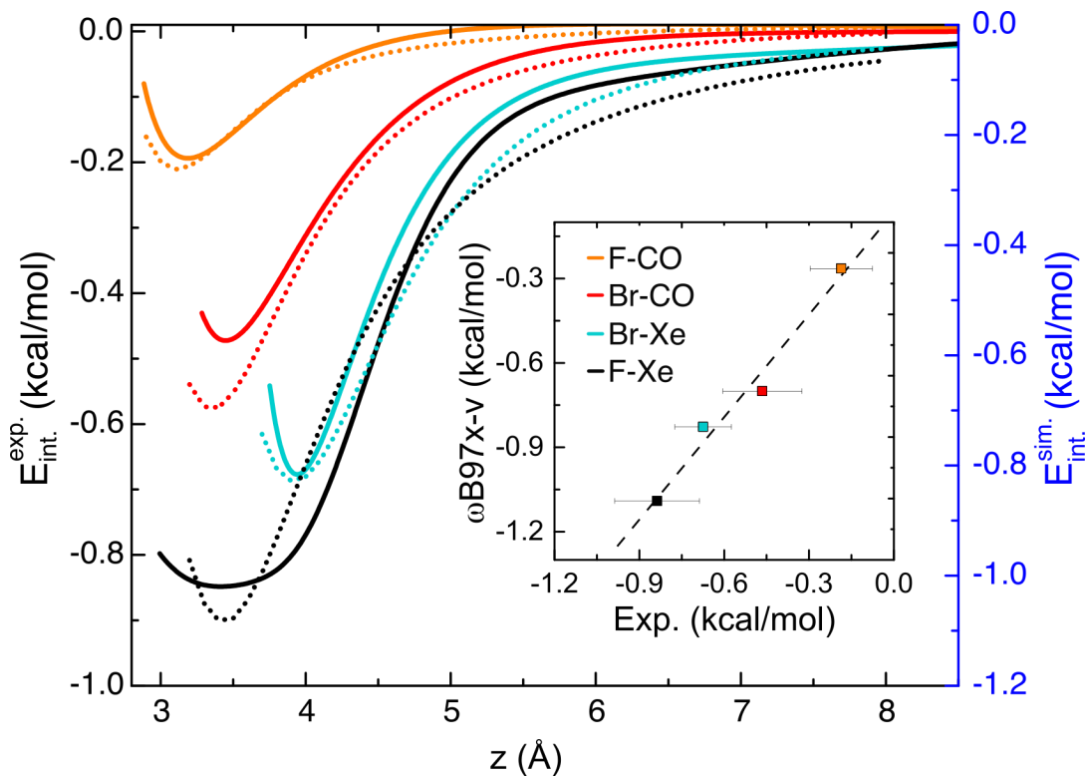

**Fig.4. Comparison of experimental and theoretical interaction energies of four complexes.** Experimental and the calculated (obtained with DFT/ωB97X-V) energy curves vs. tip-sample distance between 4FPhM and 4BrPhM molecules and Xe- and CO-tips. Inset figure displays the correlation between experimental and theoretical values ($R^2 = 0.98$) of the energy minima for all complexes; bars indicate an estimated experimental error of the energy minima.

## Supplementary Materials

## Materials and Methods

*Experimental methods*
All the experiments were conducted at $4.2\ K$ in a commercial ultra-high vacuum (UHV) system with base pressure below $5 \times 10^{-10}\ mbar$ hosting a STM/nc-AFM microscope (Createc GmbH). STM/nc-AFM images were taken with sharpened by focus ion beam (FIB) Pt/Ir tips. The tip was further cleaned and shaped in UHV conditions by gently indentation ($\sim 1\ nm$) in the bare metallic substrate. STM topographs were acquired in constant current mode with the bias voltage applied to the sample. In nc-AFM imaging, a qPlus sensor (*42*) (resonant frequency $f_o \approx 30\ kHz$; stiffness $k \approx 1800\ N/m$) was operated in frequency modulation mode keeping the oscillation amplitude constant at $50\ pm$. Both nc-AFM and KPFM images were captured in constant height mode after thermal stabilization of the tip-sample gap. The STM/nc-AFM/KPFM images were processed using WSxM software(*43*). Point force spectroscopies were obtained on top of the atoms and out of the atoms for a background reference in order to obtain the short-range interaction. The experimental interaction energy was obtained by integration of the measured force with the Sader-Jarvis method (*44*, *45*). The experimental errors were estimated respect to a polynomial fit of several spectroscopies, see **Figure S9**.

The Ag(111) substrate was prepared by repeated cycles of $Ar^+$ sputtering ($1\ keV$) and subsequent annealing at $\sim 800\ K$ in UHV.

4-bromophenyl methane and 4-fluorophenyl methane compounds were deposited by organic molecular-beam epitaxy from two different tantalum pocket maintained at $450\ K$ onto freshly clean Ag(111) samples held at room temperature in ultra-high vacuum.

*Acquisition and analysis of KPFM maps*. Experimental KPFM images representing LCPD maps were obtained from parabolic fitting of $\Delta f(V, x, y)$ data. We recorded frequency shift vs. bias in square grids of $36 \times 36$ to $48 \times 48$ points over the Br and F atoms at constant height. The acquisition time of each $\Delta f(V, x, y)$ data was of about 3 to 5 seconds. The data did not show apparent distortion or deviation from the expected parabolic behavior in the range of $[-600, 800]\ mV$ for any of the molecules (i.e., fluor or bromide tetraphenyl) and tips considered (i.e. Xe and CO-tips.). The height of acquisition was determined by point frequency shift spectroscopy to be close the well which determine the onset of repulsive interactions.

Following the acquisition of the $\Delta f(V, x, y)$ experiment, we fitted the data with a parabolic expression $\Delta f(V) = a \times (V - V_{LCPD}) + c^2$. From the fitted parabolas, we extracted the local contact potential difference $V_{LCPD}$ and plotted it in $(x, y)$ grids to generate the maps.

*Tip functionalization with CO and Xe*. CO molecules or Xe atoms were dosed on a freshly prepared molecule decorated sample at cryogenic temperatures ($< 15\ K$) in different experimental sessions. Both CO and Xe can be further removed from the metallic substrate by soft annealing the sample up to $100\ K$ for few seconds out of the cryogenic shields.

CO molecules are found individually on Ag(111) clean areas while Xe atoms condensate around 4-Br/F phenyl methane island edges and Ag step edges. For both CO and Xe apex functionalization, the tip was vertically approached towards the particle from a specific heigh with low bias voltage ($5-10\ mV$) until a characteristic jump in the current/frequency-shift is monitored.

Computational methods

*KPFM simulations*. The necessary charge densities and Hartree potentials for the KPFM simulation were calculated with VASP package(*46*, *47*), calculating separately F and Br terminated molecules in gas phase as shown in Fig 1 A,B. For tip systems, we used a Xe atom on the apex of a 10 atoms pyramid of Ag(111) and a CO molecule attached to a silver atom. For this, each structure was relaxed separately with the FHI-AIMS(*48*) package using light wave functions at GGA-PBE (*47*) level. It includes the van der Waals interactions included by the Tkatchenko–Scheffler (*49*) treatment and an energy and force convergence tolerance of $10^{-6} eV$ and $0.01\ eV/\text{Å}$ respectively. Then the electronic structure of each system was calculated with VASP carrying a self-consistency cycle under PBE approximation with a cutoff energy of $600\ eV$ and an energy tolerance of $10^{-6} eV$

The structures of the molecules adsorbed on the substrate shown in the inset of **Figure S1** were also calculated using AIMS with the parameters indicated previously. The substrate was modeled by using a 3-layer slab of 192 Ag atoms forming a $8 \times 8$ supercell.

The AFM simulated images were calculated with the probe particle model (*26*) using a stiffness of the probe apex of $0.25\ N/m$. The model includes electrostatic interaction between tip and sample from DFT calculated electronic structure of the tip and sample.

*Benchmark interaction energies.* Subsystem structures were optimized at the DFT/PBE0-D3(*40*, *41*)/cc-pVTZ level, while structures of all complexes were determined using potential energy scan where intermolecular and intramolecular distances were optimized at the ωB97X-V(*39*)/aug-cc-pVTZ and PBE0-D3/cc-pVTZ levels, respectively.

We selected 10-paramerter, range-separated hybrid GGA ωB97X-V functional with nonlocal correlation, parametrized for noncovalent interactions, since it provides excellent results for various types of noncovalent interactions. The respective RMSD is considerably better than that for other popular functionals (*50*).

Intrinsic interaction energies, $\Delta E^{INTRINSIC}$, were determined as a difference between ωB97X-V/aug-cc-pVTZ energy of a complex and energies of subsystems which were kept rigid and were optimized for isolated subsystems. Interaction energies of the model complexes were also calculated at the CCSD(T)/aug-cc-pVTZ level. Basis set superposition error was considered.

All optimization and energy calculations were performed by TURBOMOLE 7.5(*51*) using cuby4 framework(*52*, *53*). Molecular electrostatic potential (MEP) was obtained by GAUSSIAN 16(*54*) program at ωB97X-D/aug-cc-pVTZ level.

*Fitting the interaction energy curves in model complexes.* Both the measured and calculated interaction energies $V_{int}$ versus tip-sample distance $z$ were fitted using the following exponential function eq. (1)

$$V_{int} = V_0 + \sum_{i=2}^{n} F_i y^i, [y = 1 - \exp\{-a(z - z_{min})\}] \quad (1),$$

Where $V_0$; $F_i (i = 2, ..., n)$, $a$ and $z_{min}$ are fitting parameters. As a matter of fact, a quantitative reproduction of the fitted data was achieved already for $n = 3$. Interestingly, similarly as in the case of the noble gas atom pairs Ar-Xe, Kr-Xe, Xe-Xe interacting with a Xe-functionalized tip (55), the shapes of the fitted functions significantly deviate from the usually used Lennard-Jones model.

**Supplementary Text**

Theory of KPFM imaging mechanism and simulations of KPFM images

Kelvin probe force microscopy (KPFM) is an experimental technique, which senses variations of electrostatic force $F^{el}$ acting between tip and sample across the surface(17) when external bias voltage is applied. In the limit of far tip-sample distance $z$, tip and sample can be considered as the parallel plate capacitor model. When tip and sample are connected by a conductor, a charge transfer occurs between them to balance the difference of their Fermi level, see **Figure S10**. Thus, this charge transfer creates an interfacial dipole between tip and sample, which causes an attractive long-range electrostatic force $F_{LR}^{el}(z, V)$ between the tip and sample. This electrostatic force can be expressed as follows:

$$F_{LR}^{el}(z, V) = \frac{1}{2} \frac{\delta C(z)}{\delta z} (V - V_{CPD})^2 \quad (2),$$

where $C(z)$ is the capacitance of the tip-sample system, $V$ is the applied bias and $V_{CPD}$ is the so-called contact potential difference, which represents a difference between the work function of the tip $\Phi_t$ and $\Phi_s$ sample, see **Figure S10**. The eq. (2) reveals a parabolic dependence of the electrostatic interaction $F_{LR}^{el}$ on applied bias $V$. The dynamic mode of KPFM technique employs an oscillating scanning probe with well-defined eigenfrequency $f_o$ and constant amplitude, where an interaction between tip and sample causes the frequency shift $\Delta f$ of oscillating probe. Thus, the frequency shift $\Delta f$ is proportional to the force $F_{LR}^{el}$ acting between tip and sample. The KPFM measurement acquires a variation of the frequency shift $\Delta f$ on the applied bias $V$ across the surface. According to eq. (2), the contact potential difference $V_{CPD}$ equals to the applied bias voltage $V$ at which the long-range electrostatic interaction $F_{LR}^{el}$ is nullified i.e., the vertex of the Kelvin parabola $\Delta f(V)$ determines a variation of the work function across the sample.

However, this interpretation of the KPFM mechanism is only correct in limits of far tip-sample distances. If the scanning probe is brought to close tip-sample distances, typical for operation of STM and nc-AFM, then wavefunctions of frontiers atoms of tip-sample become to overlap. Consequently, an additional electrostatic interaction between atomic scale charges on tip and sample takes place. This short-range electrostatic interaction $F_{SR}^{el}$ mimics the electrostatic interaction between atomic charge densities $\rho(r, V)$ on frontiers atoms of tip and sample in external bias voltage $V$. This can give rise to atomic scale variation of the contact potential difference $V_{CPD}$ enabling atomic resolution in KPFM technique (*20, 21*). To reflect this nuance, we often use instead of $V_{CPD}$ a modified term, the local variation of the contact potential difference $V_{LCPD}$.

In the present model, we decompose the total electrostatic force $F_{tot}^{el}$ acting between tip and sample into two components:

$$F_{tot}^{el}(z,V) = F_{SR}^{el}(z,V) + F_{LR}^{el}(z,V) \quad (3).$$

The long-range electrostatic force $F_{LR}^{el}(z,V)$ is defined by eq. (2). The contact potential difference $V_{CPD}$ is considered as macroscopic quantity, which is determined by the difference between the work function of the tip $\Phi_t$ and $\Phi_s$ sample. Therefore, in our model $V_{CPD}$ does not depend on tip-sample distance $z$ and it is considered as a parameter.

In the short-range tip-sample distances in range of a few Angstroms, the short-range electrostatic interaction $F_{SR}^{el}$ will be sensitive to atomic charges on frontier atoms of both tip and sample. This makes the value of the bias $V$ that nullifies the electrostatic interaction $F_{tot}^{el}(z,V)$ sensible to the tip-sample relative position. This part of the interaction can be described by the coulombic force between charges $\rho(r,V)$ the tip and the sample, under a given applied bias and at a given tip position $\vec{z}$:

$$F_{SR}^{el}(\vec{z},V) = \iint \frac{\rho_t(r,V;\vec{z})\rho_s(r',V)}{|r-r'|^2} dr dr'. \quad (4)$$

Where $\rho_t(r,V;z)$ and $\rho_s(r',V)$ are the charge densities of the tip and sample, respectively, including the polarization induced by the applied bias $V$. For a better understanding of the system described and the interaction between them, we separate these charges in two terms, a static charge $\rho^0(r)$ and a polarized charge induced by applied bias $\delta\rho(r,V)$.

$$\rho(r,V) = \rho^0(r) + \delta\rho(r,V). \quad (5)$$

Using this form (5), we can rewrite $F_{SR}^{el}$ in eq. (4) in four terms (for simplicity we will omit variable $\vec{z}$ in $\rho_t$ for the rest):

$$F_{SR}^{el}(z,V) = \iint \frac{\rho_t^0(r)\rho_s^0(r')}{|r-r'|^2} dr dr' + \iint \frac{\delta\rho_t(r,V)\rho_s^0(r')}{|r-r'|^2} dr dr' + \iint \frac{\rho_t^0(r)\delta\rho_s(r',V)}{|r-r'|^2} dr dr' + \iint \frac{\delta\rho_t(r,V)\delta\rho_s(r',V)}{|r-r'|^2} dr dr'. \quad (6)$$

The first term of the right-hand side of eq. (6), represents the interaction between the static charges of the tip $\rho_t^0$ and the sample $\rho_s^0$, which is independent of applied bias $V$. The second term describes the interaction between the static charge density of the sample $\rho_s^0$ with the polarized charge (induced dipoles) on the tip $\delta\rho_t$. The third term accounts for the interaction between the static charge density of the tip $\rho_t^0$ with the polarized charge of the sample $\delta\rho_s$ and the last term includes interaction between polarized charges of tip $\delta\rho_t$ and sample $\delta\rho_s$.

In principle, we assume that the polarized charges $\delta\rho$ are much smaller than the static charges $\rho^0$ in ranges of applied biases $V$ typically applied in the KPFM experiment. Thus, the last term is much smaller than the other terms in eq. (6) and we neglect it in further discussion.

Let us now discuss the effect of the short-range electrostatic interaction $F_{SR}^{el}$ on the Kelvin signal $\Delta f(V)$ and resulting (sub)atomic scale contrast in KPFM. As we discussed the quadratic character of the Kelvin signal $\Delta f(V)$ is determined by the long-range electrostatic interaction $F_{LR}^{el}(z,V)$ defined by eq. (2), where the electrostatic interaction is fully compensated when the applied bias $V$ is equal to $V_{CPD}$ value. Therefore, in far tip-sample distances KPFM maps variations of the contact potential difference on mesoscopic scale without atomic resolution. However, in close tip-sample distances, the short-range electrostatic interaction $F_{SR}^{el}$ (eq, 6) driven by electrostatic interaction between atomic charges gives rise to additional correction to the Kelvin parabola $\Delta f(V)$. Thus, the precise value of the applied bias $V_{LCPD}$ required to nullify the electrostatic interaction $F_{tot}^{el}(z,V)$ will change accordingly.

Note that in eq. (6) the dependence of electrostatic interaction $F_{SR}^{el}$ on applied bias is determined only by the polarized charges $\delta\rho(V)$. For typical ranges of the applied bias adopted in the KPFM measurements, we can linearize these polarized charges in bias, i.e., $\delta\rho(r,V) = \alpha(r)V$, where $\alpha(r)$ is atomic polarizability. Consequently, we can see that second and third term can be approximated as a linear function of applied bias $V$. Therefore, these two terms cause a linear shift of the Kelvin parabola $\Delta f(V)$ defined by eq. (2). For clarity, let's take a look at an example of how the second term in eq. (6) affects the KPFM $\Delta f(V)$ signal. We consider a common wiring SPM setup, where at positive bias voltages $V > 0$ the electrons tunnel from tip to sample, see **Figure S11** b). The applied external field $\vec{E}$ then causes an induced charge $\delta\rho_t$ on tip with the negative part pointing towards the sample, as shown in **Figure S12**. In the opposite bias polarity, the orientation of the polarized charges changes accordingly. Therefore, when the KPFM probe inspects an atom or molecule on surface with a positive permanent charge $\rho_s^0$, at the positive bias voltage $V > 0$ the short-range electrostatic interaction $F_{SR}^{el}$ becomes more attractive while at the negative bias voltage $V < 0$ is repulsive. Therefore, the second term in eq. (2) gives rise to a linear correction of $\Delta f(V) \sim bV$ with a negative slope $b$, as depicted schematically in **Figure S12**, shifting the $V_{LCPD}$ towards lower values of the applied bias. In similar way, we can analyze a case of negatively charged atomic entity on the surface (**Figure S13**) as well as the contribution of the third term including interaction between permanent charge of tip $\rho_t^0$ with the induced charge $\delta\rho_t$ on sample, see **Figures S14** and **S15**. In summary, we found positive charges in the sample to shift $V_{LCPD}$ to lower values and negative charges to shift $V_{LCPD}$ to higher values (**Figure S12** and **S13**). While for charges in the tip the behavior is exactly the opposite, positive charges shift $V_{LCPD}$ to higher values (**Figure S14**) and negative charges localized on the tip apex shifts $V_{LCPD}$ to lower values (**Figure S15**).

As this short-range electrostatic interaction $F_{SR}^{el}$ is explicitly driven by atomic scale charges, it enables to achieve atomic scale resolution. Moreover, in extreme cases, when the atomic charges have strong subatomic variation, such as σ-hole, the KPFM signal can also reach subatomic resolution, as we demonstrate in the main text.

We implemented the above-described procedure for simulation of KPFM images into a home-built PP-SPM simulation tool package(26). We employ eq. (2) to obtain the parabolic character of $\Delta f(V)$. In this work, to facilitate comparison to the experimental data, we determine parameters $V_{CPD}$ and $\frac{\partial C}{\partial z}$ in eq. (2) from the experimental data. In general case, we can express the quadratic term of the force by this expression (56):

$$F_{SR}^{el}(z_{tip}, V) = \pi\varepsilon_0 \left[\frac{R_{tip}^2}{z_{tip}(z_{tip} + R_{tip})}\right](V - V_{CPD})^2. \quad (7)$$

Where $\varepsilon_0$ is the vacuum electric permeability, $R_{tip}$ is the radius of metallic tip apex and $z_{tip}$ is the tip-sample distance. First, to determine the value of the parameter $V_{CPD}$, we observed the experimental values of $V_{CPD}$ at large tip-sample distances where its value reaches the mesoscopic value, related with the difference between the work function of tip and sample and use it equation (7). Then $R_{tip}$ was adjusted so equation 7 reflects the quadratic part of the experimental interaction. For this we fit the equation $\Delta f(V, z_{tip}) = a(z_{tip}) \cdot [V - b(z_{tip})]^2 + c(z_{tip})$ to a set of simulated and experimental LCPD parabolas at different heights and choose $R_{tip}$ to obtain similar behavior with $z_{tip}$ of $a(z_{tip})$ for the simulated and experimental data. The values of these parameters for each studied system are collected in **Table S1**.

|  | F-Xe | Br-Xe | F-CO |
|---|---|---|---|
| $V_{CDP}$ (mV) | 210 | 75 | -100 |
| $R_{tip}$ (nm) | 8 | 8 | 4 |

**Table S1. Parameterized values used in simulated KPFM images**. Values of tip radius $R_{tip}$ and the contact potential difference $V_{cpd}$ adjusted for the case of F-Xe and Br-Xe: F and Br-terminated molecule measured with Xe-tip and F-CO: F-terminated molecule measured with CO-tip.

To this parabolic $\Delta f(V)$ curve we add the second and third term in eq. (6) representing the linear shift due to interaction between the static charge of tip $\rho_t^0$ and the induced charge of the sample $\delta\rho_s$ and vice versa. First, we carry out total energy DFT simulations to obtain fully optimized structure of tip and sample and corresponding Hartree potential $V_H$ and static charges $\rho^0$. On top of this, we calculate the reference charges $\rho(E_{ref})$ of both the tip and sample in an external reference field $E_{ref} = -0.1\ eV/\text{Å}$ performing self-consistent DFT calculations with the fixed optimized structure. We assume that the polarizability of tip and sample remains constant within a typical range of applied bias in experiments. Thus, we can use the linear extrapolation of the polarized charge under any given electric field in our range from the polarization at one given bias calculated with DFT:

$$\delta\rho(E) = \frac{E}{E_{ref}} \cdot [\rho(E_{ref}) - \rho^0] = \frac{E}{E_{ref}} \cdot \delta\rho(E_{ref}),$$

Where the electric field $E$ induced by applied bias $V$ can be approximated as $E = V/z_{tip}$ to determine the scaling factor to obtain the polarized charge $\delta\rho(E)$ at every given bias $V$ and tip-sample distance $z_{tip}$.

After that, we have all variables to evaluate the short-range electrostatic interaction $F_{SR}^{el}$ acting between tip and sample at given bias voltage and tip-sample position using FFT algorithm and its conversion to the frequency shift following standard procedures implemented in the PP-AFM code. Finally, we estimate value of the local contact potential difference $V_{LCPD}$ to obtain a corresponding KPFM image at a given tip position.

Structures and energies

*Isolated systems*. MEP of isolated tetra (para-X-phenyl) methane (X=F, Br) as well as of Y-Ag-tip (Y= Xe, CO) are shown in **Figure S16**. Evidently, systems with Br exhibit a significant positive σ-hole (see **Table S2**) while systems with F show uniform distribution of electron density around F atom which is negative. Further, Xe at the Ag-tip is positive contrary to O(CO) which is negative (**Table S2**). Notice that CO molecule, bound to Ag-tip via C, has negative oxygen. In an isolated CO molecule oxygen is positive. Consequently, we expect attractive and repulsive electrostatic interactions for F- and Br-systems with Xe-tip and repulsive and attractive electrostatic interactions for F- and Br-systems with CO-tip. **Figure S17** and **Table S2** present structures of model subsystems (fluoro- and bromobenzene) and their characteristics. Evidently, characteristics of model and real systems are similar what justify use of the former ones for benchmarking purposes.

| *Molecule/studied atom* | $V_{smax}$ *(kcal/mol)* |
|:---:|:---:|
| $C(C_6H_4F)_4/F$ | -12.37 |
| $C(C_6H_4Br)_4/Br$ | 15.21 |
| $XeAg_6/Xe$ | 16.04 |
| $COAg_6/O$ | -5.08 |
| $C_6H_5F/F$ | -15.74 |
| $C_6H_5Br/Br$ | 12.40 |
| $XeAg_2/Xe$ | 20.01 |

**Table S2. $V_{smax}$ on the top of selected atoms in isolated subsystems.** Analysis of MEP in all studied subsystems. There is positive σ-hole in systems with bromine, negative surface on fluorine and oxygen of CO tip. In both tips positively polarized xenon was found.

| F-Xe(Å) | CCSD(T) | B3LYP-D3 | PBE0-D3 | B97-D | M06-2X |
|---|---|---|---|---|---|
| 3.178 | -1.053 | -0.862 | -0.870 | -0.321 | -0.670 |
| 3.321 | -1.259 | -1.091 | -1.107 | -0.735 | -0.841 |
| 3.517 | -1.285 | -1.143 | -1.178 | -0.980 | -0.887 |
| 3.721 | -1.178 | -1.054 | -1.107 | -1.025 | -0.849 |
| 5.500 | -0.350 | -0.315 | -0.320 | -0.329 | -0.284 |
| F-Xe(Å) | B2PLYP | CAM-B3LYP | ωB97X-D | ωB97m-V | ωB97X-V |
| 3.178 | 0.772 | -0.355 | -0.370 | -0.969 | -0.950 |
| 3.321 | 0.382 | -0.606 | -0.695 | -1.154 | -1.167 |
| 3.517 | 0.105 | -0.700 | -0.891 | -1.168 | -1.202 |
| 3.721 | -0.030 | -0.666 | -0.937 | -1.063 | -1.098 |
| 5.500 | -0.184 | -0.243 | -0.349 | -0.319 | -0.314 |
| Br-Xe(Å) | CCSD(T) | B3LYP-D3 | PBE0-D3 | B97-D | M06-2X |
| 3.566 | -0.656 | -0.668 | -0.606 | -0.307 | -0.371 |
| 3.708 | -0.951 | -0.762 | -0.897 | -0.762 | -0.569 |
| 4.014 | -1.039 | -1.110 | -1.037 | -1.110 | -0.539 |
| 4.332 | -0.864 | -1.062 | -0.910 | -1.062 | -0.434 |
| 5.500 | -0.330 | -0.436 | -0.351 | -0.436 | -0.228 |
| Br-Xe(Å) | B2PLYP | CAM-B3LYP | ωB97X-D | ωB97m-V | ωB97X-V |
| 3.566 | 0.242 | 0.632 | 0.093 | -0.468 | -0.445 |
| 3.708 | -0.133 | 0.213 | -0.316 | -0.740 | -0.760 |
| 4.014 | -0.389 | -0.119 | -0.638 | -0.813 | -0.886 |
| 4.332 | -0.363 | -0.156 | -0.670 | -0.673 | -0.738 |
| 5.500 | -0.163 | -0.088 | -0.341 | -0.275 | -0.266 |
| Δ(F-Br) | CCSD(T) | B3LYP-D3 | PBE0-D3 | B97-D | M06-2X |
| minimum | -0.247 | 0.006 | -0.141 | 0.131 | -0.348 |
| Δ(F-Br) | B2PLYP | CAM-B3LYP | ωB97X-D | ωB97m-V | ωB97X-V |
| minimum | 0.493 | -0.582 | -0.253 | -0.355 | -0.316 |

**Table S3. Benchmark intrinsic interaction energies (in kcal/mol) of model complexes.** $\Delta E^{INTRINSIC}$ energies were obtained at CCSD(T)/aug-cc-pVTZ level and various DFT functionals in aug-cc-pVTZ basis set. Distances in model complexes are optimized at CCSD(T)/aug-cc-pVTZ level, while subsystem geometries are relaxed at MP2/aug-cc-pVTZ level. Δ(F-Br) is defined as a difference between $\Delta E^{INTRINSIC}$ in PES minimum of F-Xe model complex and Br-Xe model complex.

*Complexes*. **Figure S17** presents structures of model complexes investigated, while **Figure S7** and **Table S3** show the benchmark CCSD(T)/aug-cc-pVTZ as well as various DFT interaction energies. The best agreement between energy minima of both complexes determined at the benchmark and DFT levels were found for ωB97X-V and PBE0-D3 functionals (0.08, 0.15 and 0.11 and 0.0 kcal/mol, respectively). Further, the benchmark difference between energy minima of both complexes is better reproduced by the former functional. Since also the interaction energies of real complexes determined with former functional agree with experimental energies better (**Figure S8** and **Figure 4**) we finally used the ωB97X-V functional. Optimization of real complexes is with this functional impractical and, thus, it was performed at the PBE0 level while energies were calculated with CPU-time demanding ωB97X-V functional.

**Supplementary Figures**

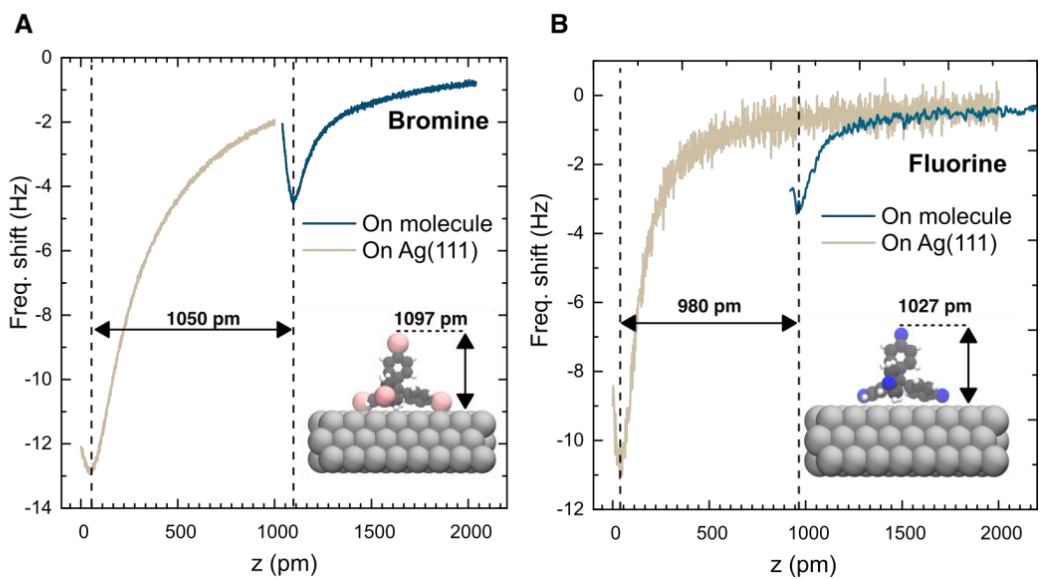

**Figure S1. Structural standing-up arrangement of 4FPhM and 4BrPhM molecules on Ag(111) surface. A-B** Experimental force spectroscopy $\Delta f(z)$ acquired with CO-tip over 4BrPhM and 4FPhM on the Ag(111) and bare Ag(111) surface. Distance between the $\Delta f(z)$ minima over the 4BrPhM and 4FPhM molecules and the bare Ag(111) surface match very well to calculated heights of the frontier atoms of 4BrPhM and 4FPhM molecules on Ag(111). This confirms the standing up configuration of the 4BrPhM and 4FPhM molecules on the Ag(111) surface.

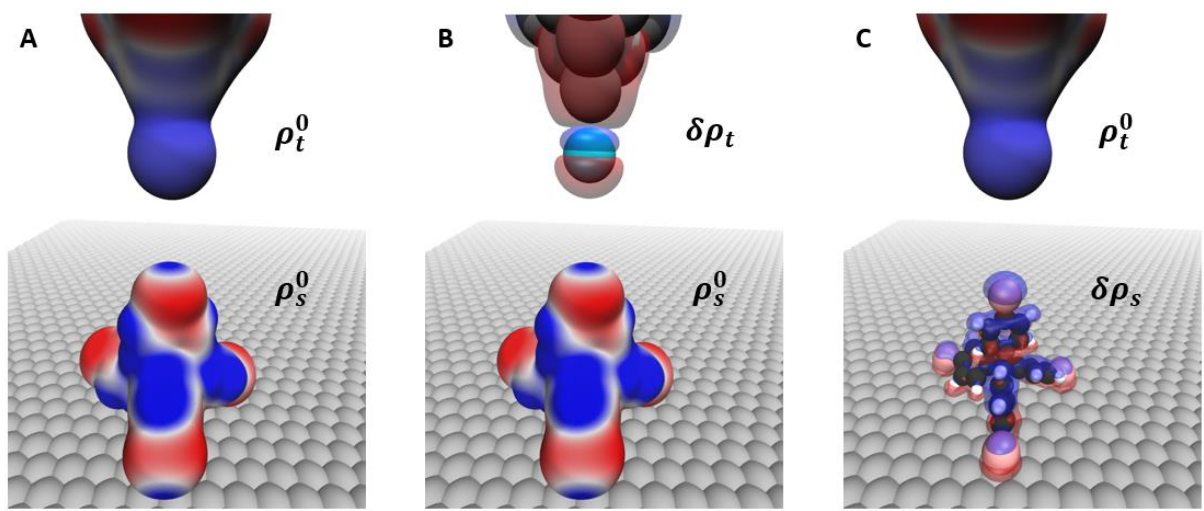

**Figure S2 Schematic description of the charge densities involved in KPFM imaging mechanism.** Here we show the three main components of the electrostatic interaction in the presence of an applied positive bias, *corresponding to the 1st(A), 2nd (B), and 3rd (C) terms of the right-hand side of equation (7)*. A) The electrostatic interaction between static charges of tip $\boldsymbol{\rho_t^o}$ and sample $\boldsymbol{\rho_s^o}$. B) The electrostatic interaction between tip polarized charge $\boldsymbol{\delta\rho_t}$ and the static charge of sample $\boldsymbol{\rho_s^o}$. C) The electrostatic interaction between the static charge of tip $\boldsymbol{\rho_t^o}$ and the polarized chare of sample $\boldsymbol{\delta\rho_s}$.

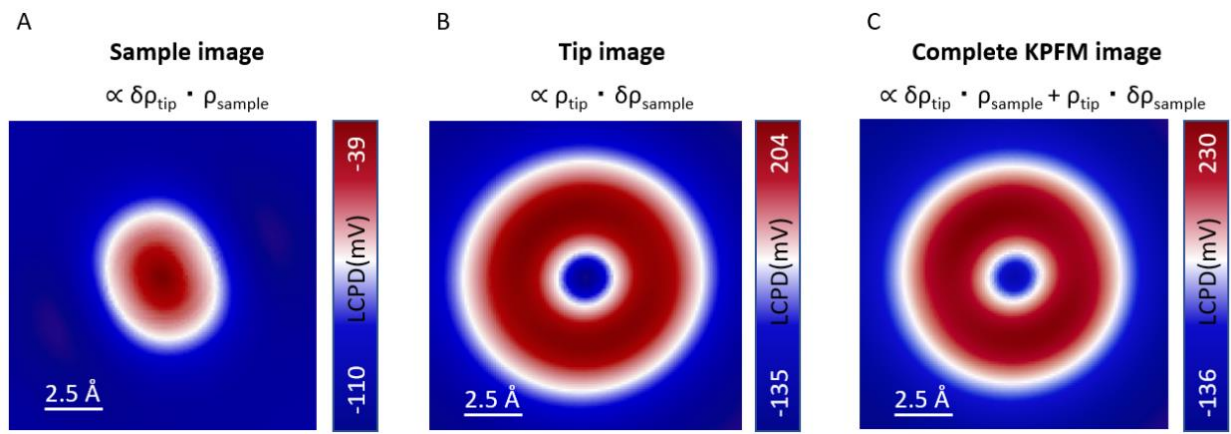

**Figure S3. Decomposition of the KPFM images of 4FPhM molecule with CO-tip.. A)** Calculated component of the KPFM image including only the interaction between the tip polarization $\delta\rho_t$ and the sample static density $\rho_s^o$. **B)** KPFM image including only the interaction between the tip static charges $\rho_t^o$ reflecting the quadrupole charge distribution on CO-tip and the spherical sample polarization $\delta\rho_s$. **C)** The total calculated KPFM image, including both components presented separately in **A)** and **B)** revealing the dominance of the second term formed by the quadrupole charge of CO molecule (see **Figure 3A**).

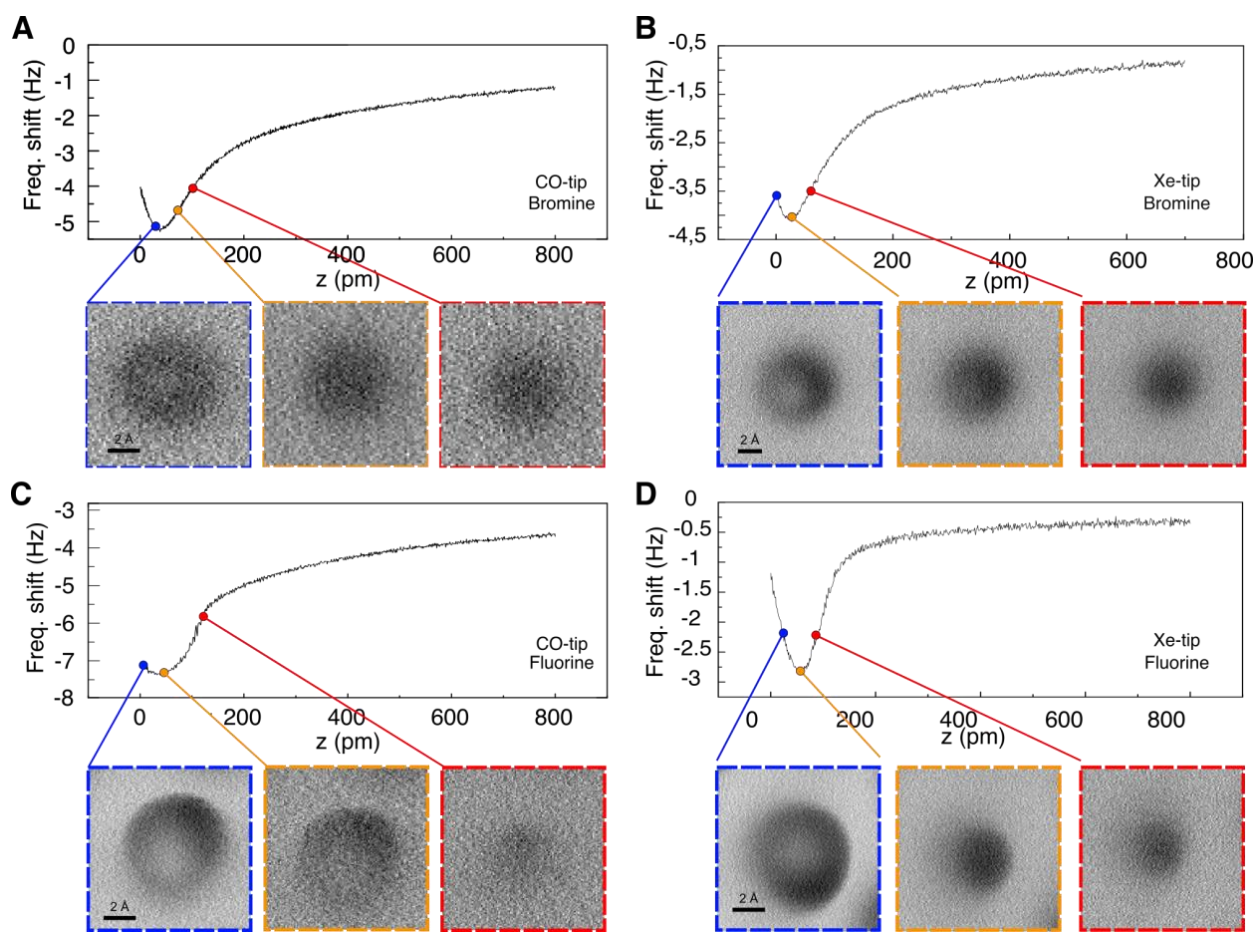

**Figure S4. Absence of subatomic contrast corresponding to σ-hole in experimental nc-AFM images with Xe and CO-tips.** Series of nc-AFM images acquired over 4BrPhM and 4FPhM molecules with Xe and CO-tips at different tip-sample heights, CO-4BrPhM (**A**), Xe-4BrPhM (**B**), CO-4FPhM (**C**), Xe-4FPhM (**D**). Upper figures show $\Delta f(z)$ spectra and lower figures display AFM images at given tip-sample distance. In close distances, after the minima in $\Delta f(z)$ spectra, a bright spherical contrast appears in all four cases, which is caused by Pauli repulsive forces. The fact that the similar contrast appears on both 4BrPhM and 4FPhM rules out influence of the σ-hole on the AFM contrast.

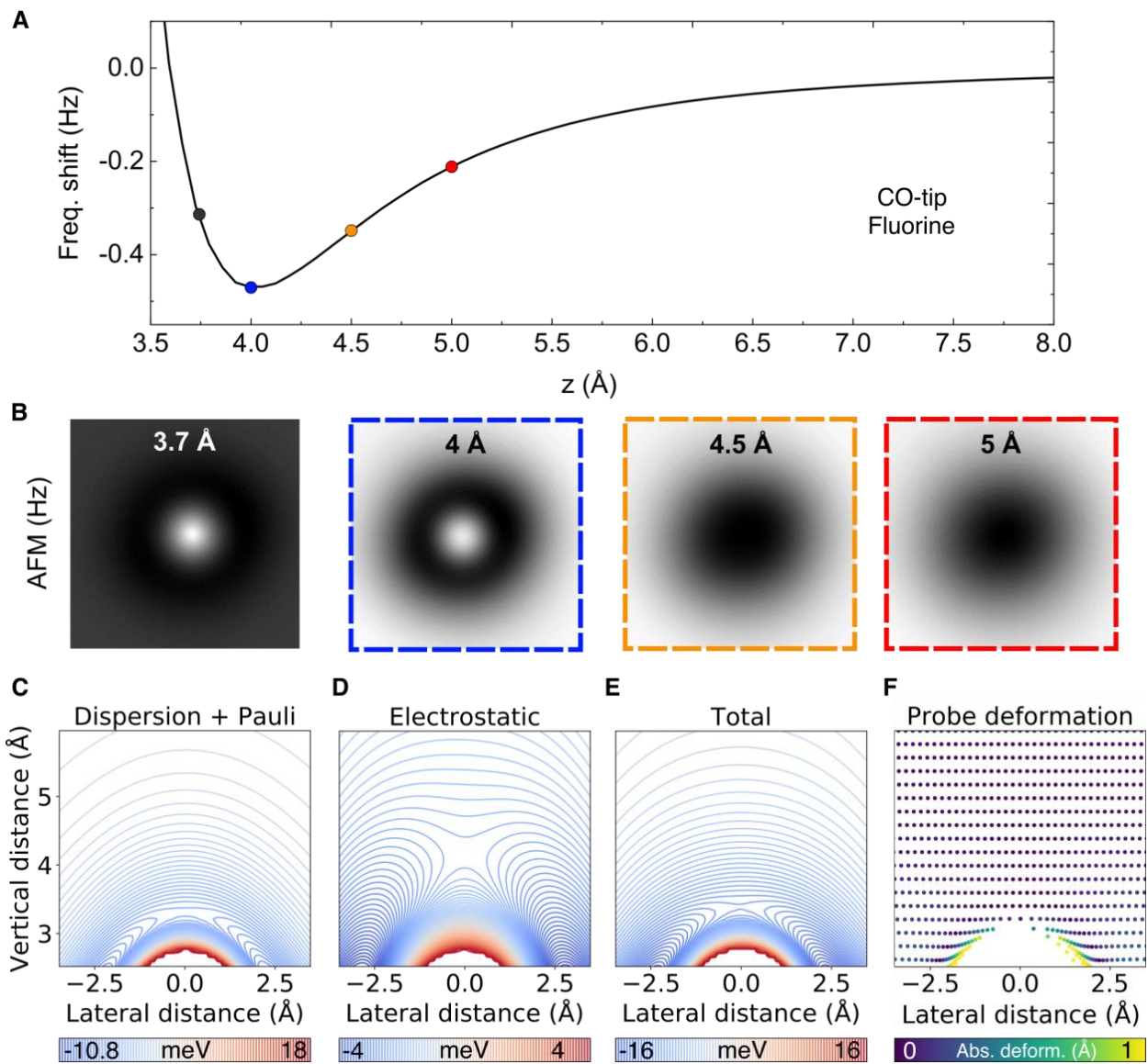

**Figure S5 Analysis of different energy components of calculated AFM images of 4FPhM molecule with CO-tip. A)** Simulated frequency shift plotted versus the vertical distance between the fluorine atom and the CO terminated apex of the probe. **B)** AFM images of fluorine terminated molecule simulated with a CO terminated tip at the indicated heights above the halogen atom, **C-E)** Lateral cross-section isoline plots over the fluorine atom of: **C)** Dispersion and Pauli repulsion interaction energy between the CO terminated tip and the fluorine atom, simulated using a Lennard-Jones potential **D)** Calculated electrostatic interaction energy between the CO terminated tip and the fluorine atom. **E)** Total interaction energy between CO terminated tip and the fluorine atom, sum of **C)** and **D)**. **F)** Lateral deformation of the tip decorating CO molecule.

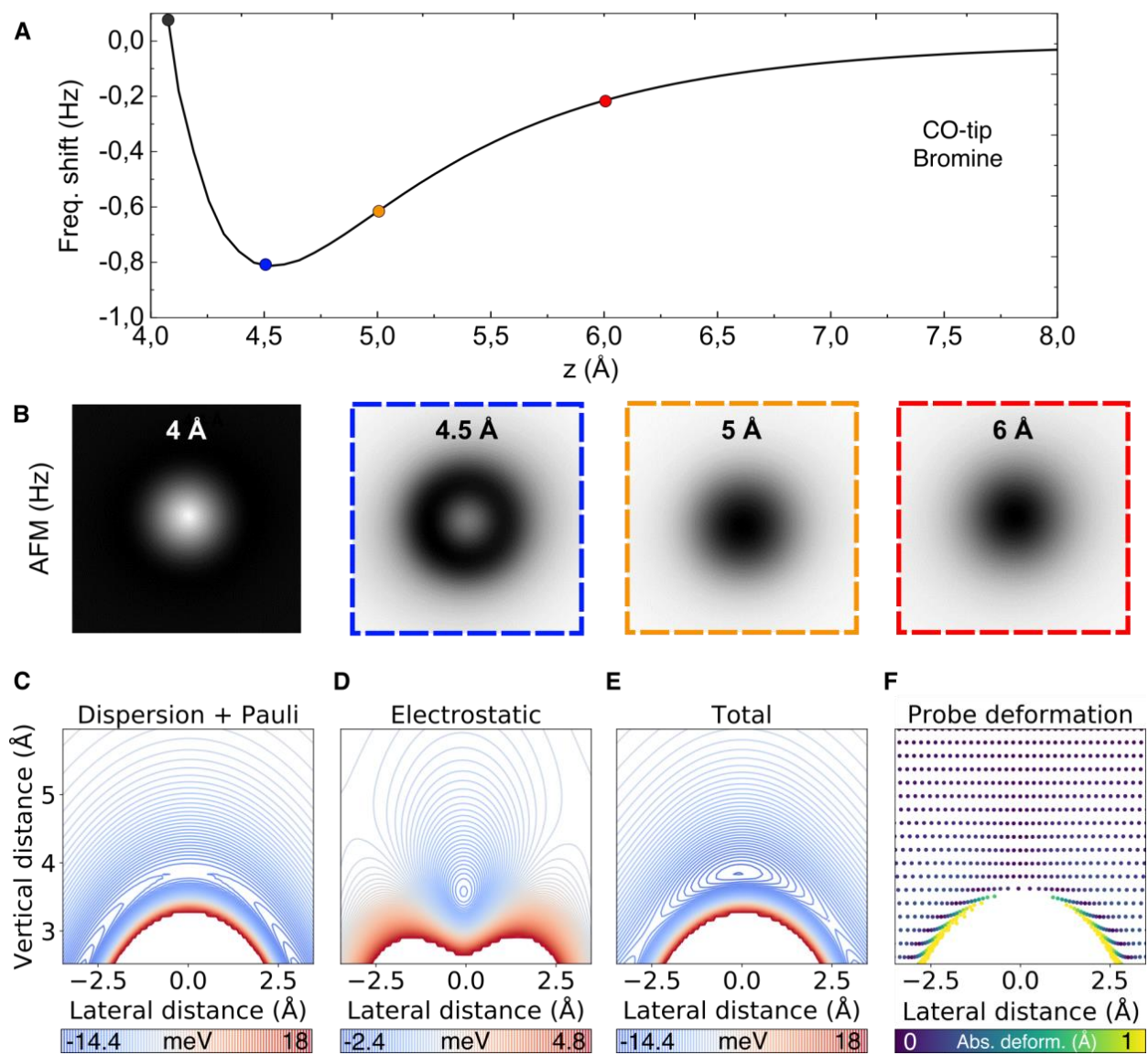

**Figure S6 Analysis of different energy components of calculated AFM images of 4BrhM molecule with CO-tip A)** Simulated frequency shift plotted versus the vertical distance between the bromine atom and the CO terminated apex of the probe. **B)** AFM images of bromine terminated molecule simulated with a a CO terminated tip at the indicated heights above the halogen atom, **C-D)** Lateral cross-section isoline plots over the bromine atom of: **C)** Dispersion and Pauli repulsion interaction energy between the CO terminated tip and the bromine atom, simulated using a Lennard-Jones potential **D)** Calculated electrostatic interaction energy between the CO terminated tip and the bromine atom. **E)** Total interaction energy between the CO terminated tip and the bromine atom, sum of c) and **E)**. **F)** Lateral deformation of the tip decorating CO molecule.

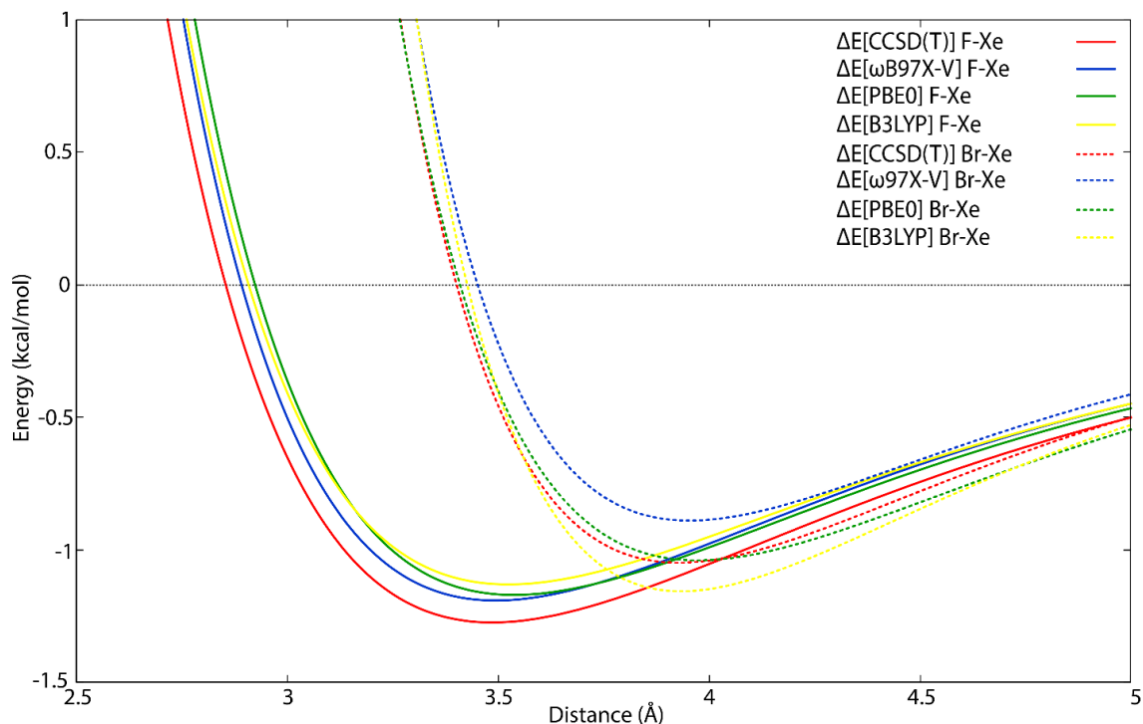

**Figure S7. Comparison of $\Delta E^{INTRINSIC}$ vs. distance curves of the model complexes.** Benchmark CCSD(T)/aug-cc-pVTZ curve (red) is compared with various DFT functionals curves [ωB97X-V (blue), PBE0 (green) and B3LYP (yellow)] for the small model complex shown in Figure S17.

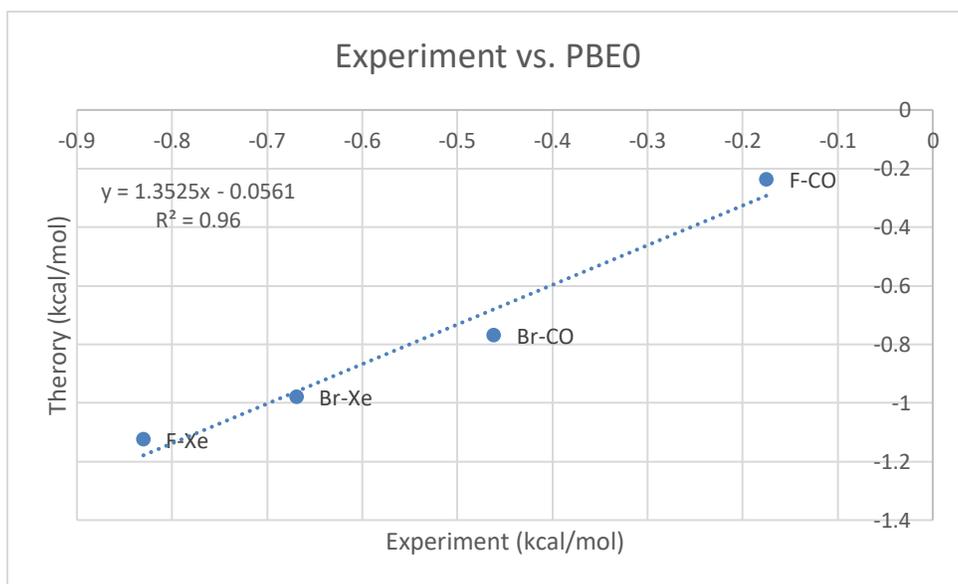

**Figure S8. Correlation between energy minima found for all four real complexes experimentally and computationally (PBE0/aug-cc-pVTZ).** Agreement between experimental and PBE0 values is significantly inferior than in the case of ωB97X-V (see **Figure 4**). The correlation coefficients ($R^2$) determined for PBE0 and ωB97X-V are equal to 0.96 and 0.98, respectively.

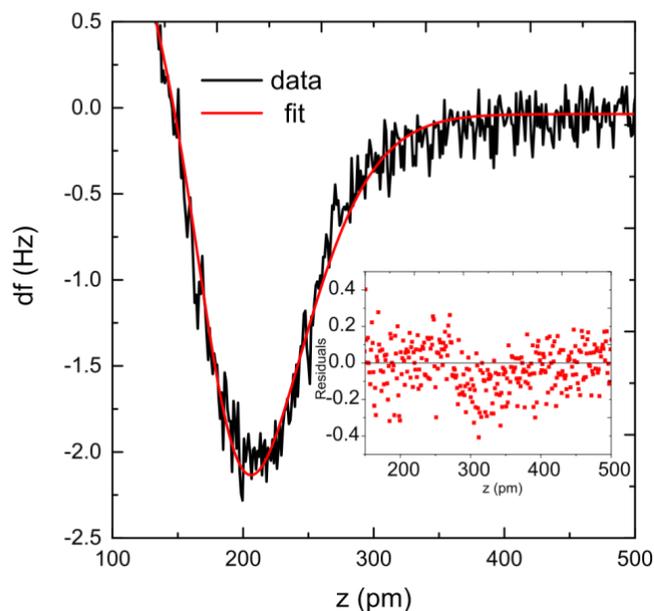

**Figure S9. Estimation of the experimental error for the case F-Xe.** The experimental error of the point spectroscopy energies was estimated by calculating the residuals of a polynomial fit. The residual is then converted into a force using the Sader-Jarvis method and integrated to obtain an energy value.

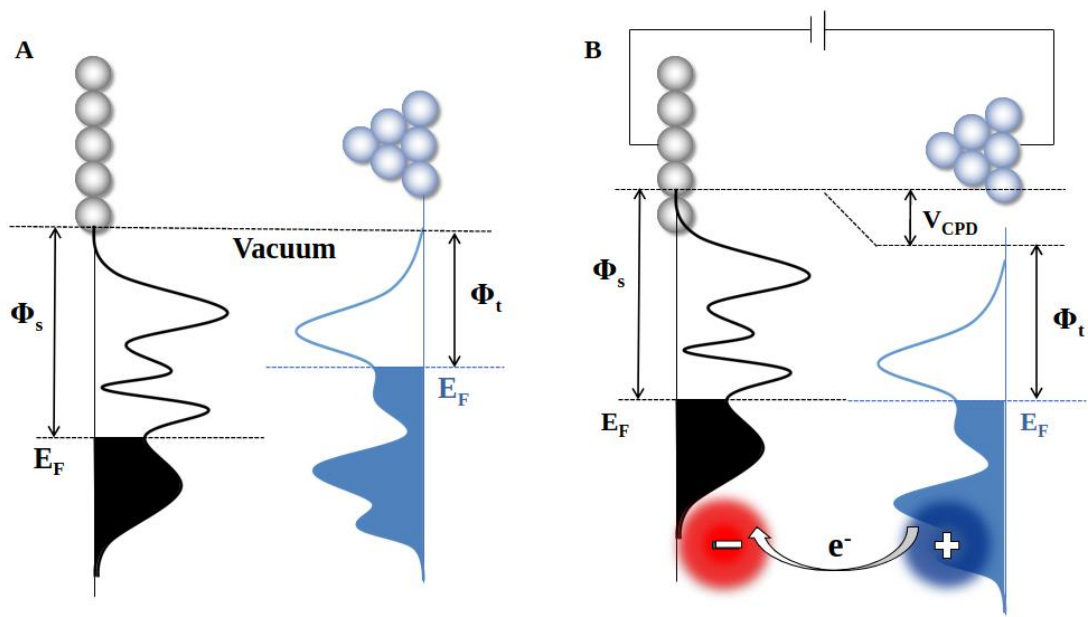

**Figure S10. Schematic view of the origin of the electrostatic interaction in KPFM. A** Alignment of the Fermi level of tip and sample, before they are connected by an electric circuit. **B**

When tip and sample are brought into contact, the difference between work functions of sample $\Phi_s$ and tip $\Phi_t$ causes charge transfer between the tip and the sample forming an additional interfacial dipole across the tunnelling junction. Compensation of this field resulting from the interfacial dipole by an external bias allows one to determine the contact potential difference $V_{CPD}$.

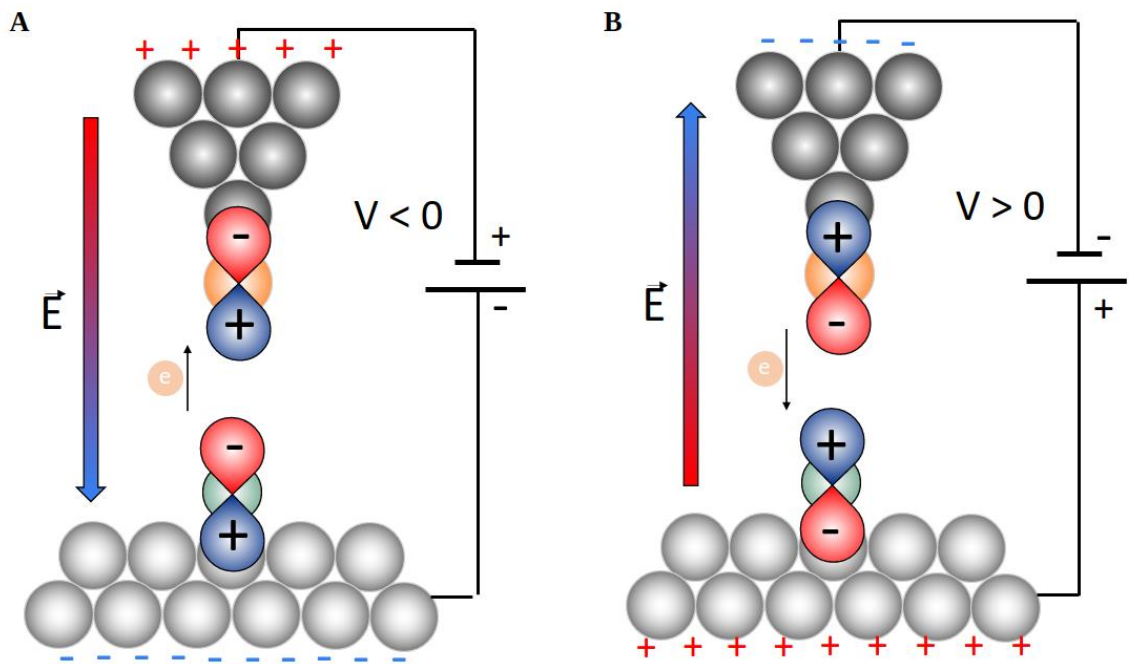

**Figure S11. Schematic sketch of charge polarization in KPFM with applied bias voltage. A** Polarization of tip and sample under a negative bias applied to the sample, creating an electric field pointing from the tip to the sample. This field polarizes negatively the topmost areas of the sample and positively the apex of the tip. **B** Polarization of tip and sample under a positive bias applied to the sample, creating an electric field pointing from the sample to the tip. This field polarizes positively the topmost areas of the sample and negatively the apex of the tip.

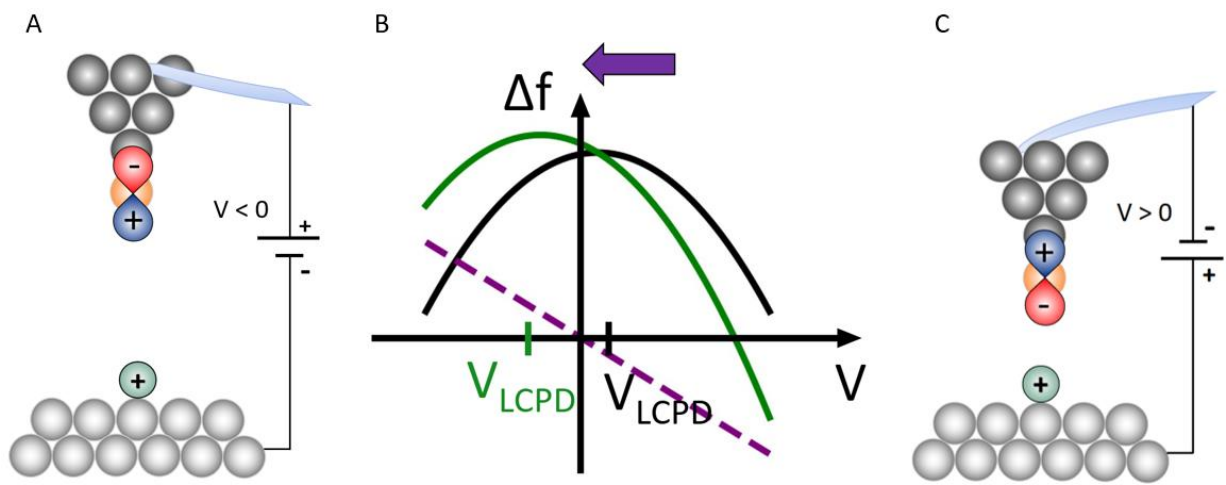

**Figure S12. Schematic sketch of variation of $V_{LCPD}$ due to short-range electrostatic interaction between the tip polarization and a localized positive charge in the sample.** Here we show the contribution of the 2nd term of equation (7) to the LCPD for the case of a positive static charge in the sample. The tip will polarize with bias according to the scheme in figure S11, resulting in repulsive contribution for V < 0 (**A**) and attractive for V > 0 (**C**). This will give this linear term a negative slope (purple dashed line in **B**), shifting the value of the mesoscopic $V_{LCPD}$ (black parabola in **B**) towards lower values ($V_{LCPD}$ of the green parabola in **B**).

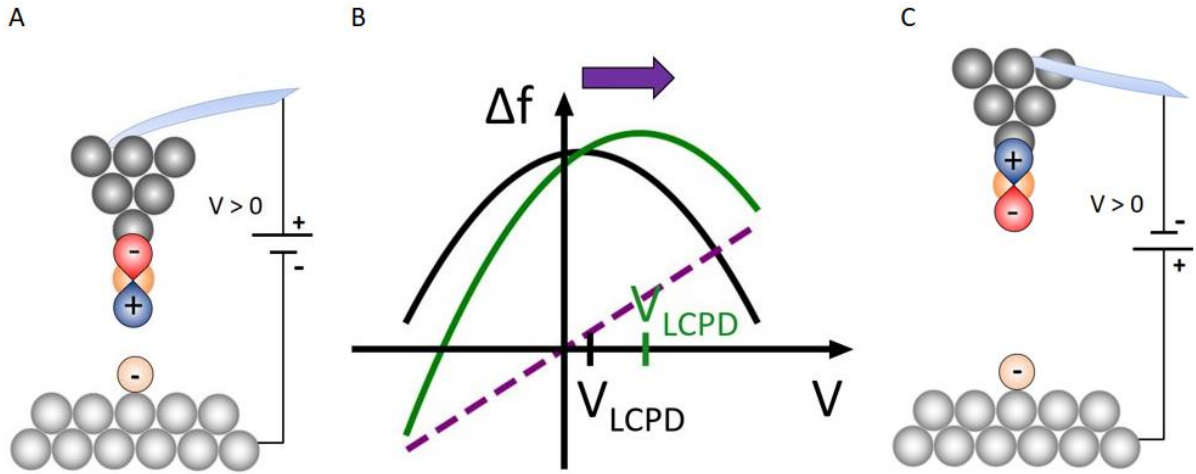

**Figure S13. Schematic sketch of variation of $V_{LCPD}$ due to short-range electrostatic interaction between the tip polarization and a localized negative charge in the sample.** Here we show the contribution of the 2nd term of equation 6 to the LCPD for the case of a negative static charge in the sample. The tip will polarize with bias according to the scheme in figure S2, resulting in attractive contribution for V < 0 (**A**) and repulsive for V > 0 (**C**). This will give this linear term a positive slope (purple dashed line in **B**), shifting the value of the mesoscopic $V_{LCPD}$ (black parabola in **B**) towards higher values ($V_{LCPD}$ of the green parabola in **B**).

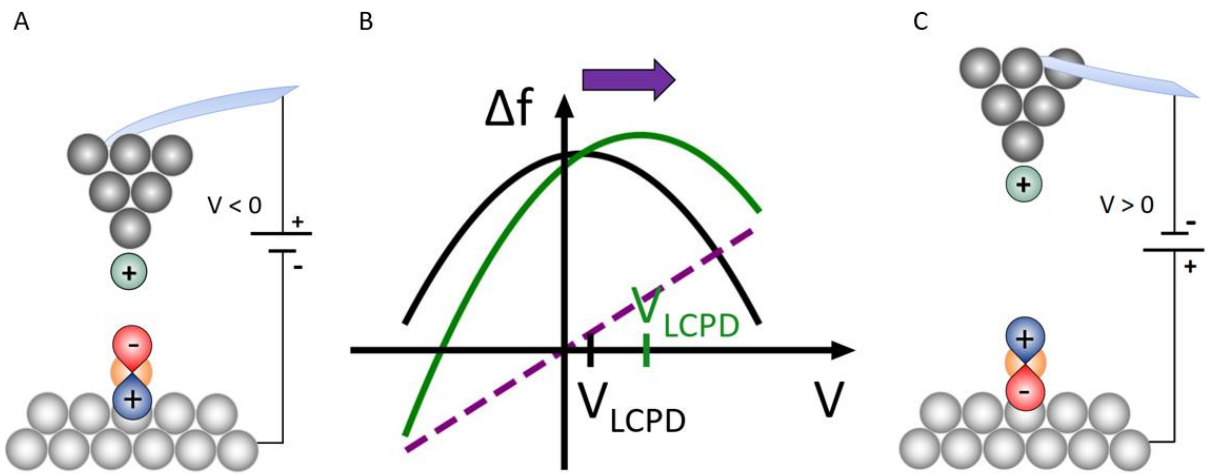

**Figure S14. Schematic sketch of variation of $V_{LCPD}$ due to short-range electrostatic interaction between the sample polarization and a localized positive charge in the tip apex.** Here we show the contribution of the 3$^{th}$ term of equation 7 to the LCPD for the case of a positive static charge in the tip. The sample will polarize with bias according to the scheme in figure S11, resulting in attractive contribution for V < 0 (**A**) and repulsive for V > 0 (**C**). This will give this linear term a positive slope (purple dashed line in **B**), shifting the value of the mesoscopic $V_{LCPD}$ (black parabola in **B**) towards higher values ($V_{LCPD}$ of the green parabola in **B**).

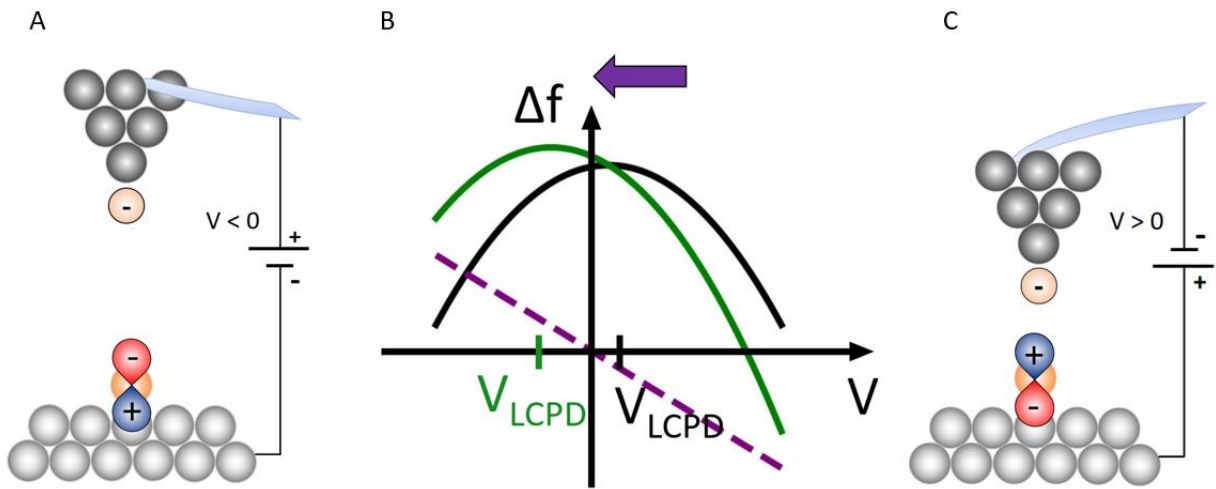

**Figure S15. Schematic sketch of variation of $V_{LCPD}$ due to short-range electrostatic interaction between the sample polarization and a localized negative charge in the tip apex.** Here we show the contribution of the 3$^{th}$ term of equation 7 to the LCPD for the case of a negative static charge in the tip. The sample will polarize with bias according to the scheme in figure S11, resulting in repulsive contribution for V < 0 (**A**) and attractive for V > 0 (**C**). This will give this linear term a negative slope (purple dashed line in **B**), shifting the value of the mesoscopic $V_{LCPD}$ (black parabola in **B**) towards lower values ($V_{LCPD}$ of the green parabola in **B**).

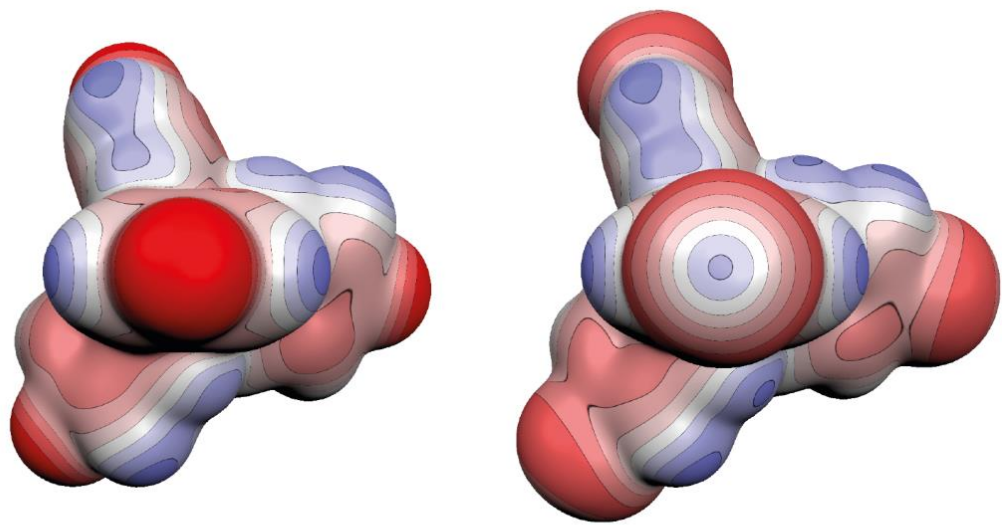

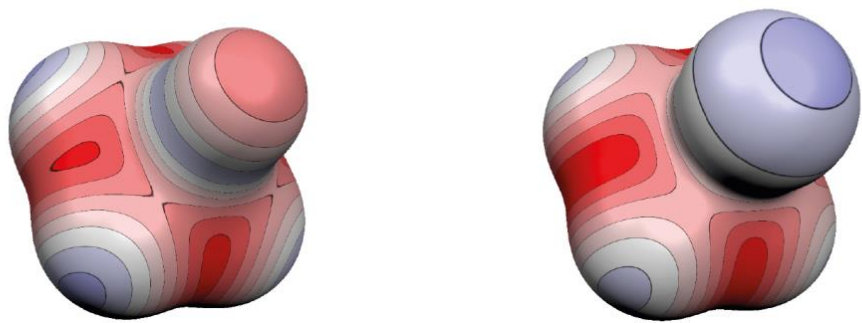

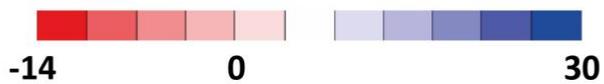

**Figure S16. MEP of real complexes.** Here we show MEP computed at ωB97X-D/aug-cc-pVTZ level, on 0.001 a.u. molecular surfaces of real subsystems studied. Color ranges are in kcal/mol.

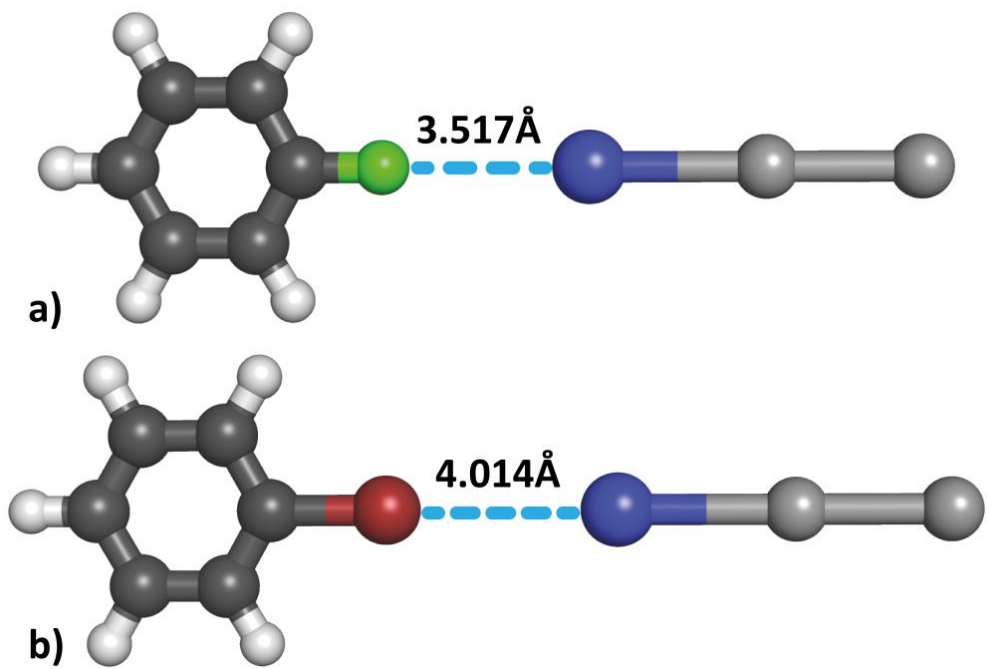

**Figure S17. Ball-stick structures of optimized model complexes.** Distances in model complexes are optimized at CCSD(T)/aug-cc-pVTZ level, while subsystem geometries are relaxed at MP2/aug-cc-pVTZ level. Structure a) show fluorobenzene interacting with Xe tip model (XeAg$_2$). Structure b) show bromobenzene with Xe tip model (XeAg$_2$). Hydrogen is shown in white, carbon in dark gray, fluorine in green, bromine in burgundy, xenon in blue and silver in light grey.